\documentclass{aa}
\usepackage[varg]{txfonts} %aa requirement
\usepackage{bm} %for vectors
\usepackage{upgreek} %for stress tensor
\usepackage{amsmath,amsfonts,amssymb} %for math
\usepackage{hyperref} %to edit citation properties
\usepackage{nicefrac} %for some nice inline fractions
\usepackage{placeins}

\hypersetup{
	pdftitle = {VSI and planets},
	pdfauthor = {Alexandros Ziampras, Wilhelm Kley, Richard P. Nelson},
	pdfsubject = {},
	colorlinks = {true},
	linkcolor = [rgb]{0,0.35,0.7},
	citecolor = [rgb]{0,0.35,0.7},
	filecolor = [rgb]{0.61,0,0},
	urlcolor = [rgb]{0.61,0,0},
}

\usepackage{graphicx}  %for images
\graphicspath{{arxiv-figures/}}

\newcommand{\tensor}[1]{\overline{\textbf{#1}}}

\newcommand{\DP}[2]{\frac{\partial{#1}}{\partial{#2}}}
\newcommand{\D}[2]{\frac{\text{d}{#1}}{\text{d}{#2}}}
\newcommand{\G}{\text{G}}
\newcommand{\Mstar}{M_\star}
\newcommand{\Rp}{R_\text{p}}
\newcommand{\Mp}{M_\text{p}}
\newcommand{\Mth}{M_\text{th}}
\newcommand{\Mearth}{\text{M}_\oplus}
\newcommand{\Msun}{\text{M}_\odot}
\newcommand{\Mjup}{\mathrm{M}_\mathrm{J}}
\newcommand{\Rgas}{\mathcal{R}}
\newcommand{\cs}{c_\text{s}}

\newcommand{\csadb}{c_\text{s}^\text{ad}}
\newcommand{\OmegaK}{\Omega_\text{K}}
\newcommand{\mean}[1]{\langle {#1} \rangle}
\newcommand{\cv}{c_\text{v}}

\newcommand{\vel}{\bm{u}}
\newcommand{\aacc}{\alpha_\text{acc}}
\newcommand{\amix}{\alpha_\text{mix}}
\newcommand{\tot}{\text{tot}}
\newcommand{\VSI}{\text{VSI}}

\begin{document}

\title{Hydrodynamic turbulence in disks with embedded planets}

\author{
    Alexandros~Ziampras\inst{\ref{inst1},\ref{inst2}},
    Wilhelm~Kley\inst{\ref{inst1}}\thanks{W.~Kley is deceased and is included as a co-author for his significant contribution and guidance over the course of this project.},
    \and
    Richard~P.~Nelson\inst{\ref{inst2}}
}

\date{}

\institute{
	Institut f{\"u}r Astronomie und Astrophysik, Universit{\"a}t T{\"u}bingen, Auf der Morgenstelle 10, 72076 T{\"u}bingen, Germany\label{inst1}
    \and
	Astronomy Unit, School of Physical and Chemical Sciences, Queen Mary University of London, London E1 4NS, UK\label{inst2}
    \\
    \email{alexandros.ziampras@uni-tuebingen.de, r.p.nelson@qmul.ac.uk}
}

\abstract
{
	The vertical shear instability (VSI) is a source of hydrodynamic turbulence that can drive vigorous vertical mixing and moderate levels of accretion in protoplanetary disks, and it could be observable in the near future. With high-resolution three-dimensional numerical hydrodynamics simulations, we modeled the behavior of the VSI in protoplanetary disks with and without embedded planets. We then measured its accretion and mixing capabilities by comparing the full Reynolds stress, which includes the contribution of nonaxisymmetric features, such as spiral arms and vortices, to the Reynolds stress due to the azimuthally averaged velocity field, which can be attributed to good approximation to the VSI. We verified that the VSI can contribute to the accretion stress and showed that, depending on disk conditions, an embedded planet can coexist with or suppress VSI turbulent stress. Specifically, the presence of spiral shocks launched by a planet or planet-generated vortices can interfere with the VSI near the planet's vicinity, with the instability recovering at large enough distances from the planet or vortex. Our results suggest that observations of VSI signatures are unlikely in disks that contain massive, nonaxisymmetric features.
}


\bibpunct{(}{)}{;}{a}{}{,}

\maketitle

\section{Introduction}
\label{sec:introduction}

%intro: turbulence in general
Turbulence is thought to be a key mechanism that affects the evolution of circumstellar disks, as well as the planet formation process. From facilitating accretion of gas onto the central star through sustained outward angular momentum transport \citep{lynden-bell-pringle-1974,shakura-sunyaev-1973,balbus-papaloizou-1999}, to the vertical stirring of dust grains observable in millimeter observations with the ALMA telescope \citep[e.g.,][]{andrews-etal-2018, dullemond-etal-2018, villenave-etal-2022}, turbulence plays a significant role in setting the environment in which planets form and grow.

%turbulence in planet-disk interactio 
In the context of planet--disk interaction, understanding turbulence is fundamentally important in several aspects. In the immediate vicinity of an embedded planet, turbulence can regulate the accretion of gas and dust particles onto the planet, influencing its growth \citep{picogna-etal-2018}. In the case of massive planets, which can strongly affect the disk background by opening gaps and forming pressure bumps in the gas \citep[e.g.,][]{kley-nelson-2012}, turbulence dictates the formation and decay of vortices at the planet's gap edge \citep[e.g.,][]{li-etal-2005,devalborro-etal-2007,rometsch-etal-2021}, as well as the radial diffusion of dust that accumulates on the resulting pressure bumps \citep{dullemond-etal-2018} that can lead to observable rings in continuum emission \citep{huang-etal-2018}. As a result, understanding how turbulence operates on both large and small scales is necessary in studying planet--disk interaction.

%why focus on the VSI
A variety of (magneto-)hydrodynamical mechanisms have been proposed as a means of generating turbulence, depending on the mass, radial, and vertical structure, as well as the ionization fraction of the disk \citep[for a comprehensive review, see][]{lyra-umurhan-2019}. Of particular interest to our work is the vertical shear instability \citep[VSI,][]{nelson-etal-2013}, which is expected to be active in the ``dead zone'' of the disk, where the magnetorotational instability \citep[MRI,][]{balbus-hawley-1991} is suppressed due to low gas ionization fractions \citep{gammie-1996} or nonideal MHD effects \citep{cui-bai-2020} and therefore it cannot operate efficiently. 

%the VSI
The VSI, originally conceived by \citet{goldreich-schubert-1967} and \citet{fricke-1968} in the context of rotating stars, and having been brought into focus in the protoplanetary disk context by \citet{nelson-etal-2013}, has received extensive attention in the last decade as it arises naturally in typical protoplanetary disks. It requires the presence of a vertical shear (i.e., a height-dependent rotation profile) and relatively short cooling timescales \citep{lin-youdin-2015}, both of which are satisfied at distances of $\gtrsim$10\,au for typical disk parameters in models with realistic disk thermodynamics \citep{stoll-kley-2014,flock-etal-2017} and up to a height of approximately 3 gas scale heights when accounting for dust--gas coupling \citep{pfeil-klahr-2021}. Its ability to produce accretion rates with a turbulent $\alpha\sim10^{-4}$--$10^{-3}$ \citep{nelson-etal-2013,stoll-kley-2014,manger-klahr-2018,flores-etal-2020} coupled with the strongly anisotropic vertical stirring it generates \citep{stoll-etal-2017b} deem the VSI a mechanism compatible with observed levels of turbulence in protoplanetary disks \citep[e.g.,][]{dullemond-etal-2018,flaherty-etal-2020}.

%planet/vortex-VSI interaction
Regarding an embedded planet's influence on the VSI, \citet{stoll-etal-2017a} have shown that VSI-active disks behave very similarly to traditionally viscous ones in terms of the gap structure sculpted by and the migration torques acting on a planet of up to $100\,\Mearth$. They have also shown that, even in this high-mass case, the disk maintains an accretion turbulence parameter $\alpha\sim10^{-4}$. The vertical stirring efficiency of the VSI, however, is not addressed quantitatively, apart from an apparent damping of the gas vertical velocity in the outer disk for $\Mp=100\,\Mearth$, attributed in part to vortex activity. This possible quenching of the VSI has important implications for any hopes of observing the kinematic signatures of the VSI in molecular line emission, as its signature vertical motion is extremely faint and very likely not observable even by modern ALMA standards in disks without an azimuthal structure \citep{barraza-etal-2021}.

%what we do
Our goal is to study the interaction between an embedded planet or vortex and the VSI and isolate the contribution of the latter, so as to quantify the extent to which the instability can coexist alongside nonaxisymmetric features such as planet-generated vortices and spirals. In doing so, we aim to gauge whether an observation of the VSI in disks with an azimuthal structure could be realized in the near future using CO kinematics with ALMA.

We describe our physical setup and numerical methods in Sect.~\ref{sec:physics-numerics}. We present our results in Sect.~\ref{sec:results}, and discuss their implications in Sect.~\ref{sec:discussion}. We finally list our conclusions in Sect.~\ref{sec:conclusions}.

\section{Physics and numerics}
\label{sec:physics-numerics}

In this section we first provide a brief rundown of the physical model and parameters we used and motivate our choices. We then describe our numerical setup and the methods used to analyze our results, highlighting the specifics of the averaging procedure for our time- and space-dependent datasets.

\subsection{Physics}
\label{sec:physics}

We solve the inviscid, time-dependent three-dimensional (3D) hydrodynamics equations \citep{tassoul-1978} for an ideal gas with density $\rho$, pressure $P$ and velocity field $\vel$ orbiting around a star with mass $M_\star=1\,\Msun$
\begin{align}
	\label{eq:hydro1}
	\DP{\rho}{t} & + \vel\cdot\nabla\rho = -\rho\nabla\cdot\vel \\\label{eq:hydro2}
	\DP{\vel}{t} & + (\vel\cdot\nabla)\vel = -\frac{1}{\rho}\nabla P -\nabla\Phi \\\label{eq:hydro3}
	\DP{e}{t} & + \vel\cdot\nabla e = -\gamma e\nabla\cdot\vel + S,
\end{align}
where $\Phi=-\G M_\star/r$ is the gravitational potential of the star at distance $r$, $e = P/(\gamma-1)$ is the thermal energy density of the gas, and $S$ encapsulates any additional terms that can affect the gas temperature $T$. The gas has an adiabatic index $\gamma=\nicefrac{7}{5}$ and a mean molecular weight $\mu=2.35$ such that $e = \rho\cv T$, with $\cv=\frac{\Rgas}{\mu(\gamma-1)}$ being its heat capacity at constant volume. The gravitational constant and gas constant are indicated with $\G$ and $\Rgas$, respectively. Finally, the isothermal sound speed of the gas is given by $\cs=\sqrt{P/\rho}$ and relates to the adiabatic sound speed through $\csadb=\sqrt{\gamma}\cs$.

We now adopt a cylindrical coordinate system $\{R,\phi,z\}$ where $r=\sqrt{R^2+z^2}$. We can derive a hydrodynamic equilibrium state for an axisymmetric ($\nicefrac{\partial}{\partial\phi}=0$), nonaccreting disk ($\vel=u_\phi\hat{\phi}$) with a vertically isothermal temperature profile ($\nicefrac{\partial T}{\partial z} = 0$). If we further assume that the gas density at the midplane $\rho_\text{mid}$ as well as the temperature both follow power law profiles in the radial direction $R$ with
\begin{equation}
	\label{eq:power-law}
	\rho_\text{mid}(R) = \rho_{z=0} = \rho_0 \left(\frac{R}{R_0}\right)^p,\qquad T(R) = T_0 \left(\frac{R}{R_0}\right)^q,
\end{equation}
we can write the density and velocity profiles in equilibrium \citep[see for example][]{nelson-etal-2013}
\begin{align}
	\label{eq:equilibrium1}
	\rho^\text{eq}(R,z) &= \rho_\text{mid}(R)\,\exp\left[\frac{\G \Mstar}{\cs^2}\left(\frac{1}{r} - \frac{1}{R}\right)\right],\\\label{eq:equilibrium2}
	u_\phi^\text{eq}(R,z) &= R\OmegaK\left[1 + (p+q)\left(\frac{H}{R}\right)^2 + q\left(1-\frac{R}{r}\right)\right]^{\nicefrac{1}{2}},
\end{align}
where $\OmegaK(R)=\sqrt{\G M_\star/R^3}$ is the Keplerian angular velocity and $H(R) = \cs/\OmegaK$ is the pressure scale height. Through the latter we can also define the aspect ratio $h=H/R$.

We choose to relax the temperature to its initial profile $T_{t=0}$ using a parametrized $\beta$-cooling approach \citep[e.g.,][]{gammie-2001}, where the cooling timescale is given by $t_\text{cool} = \beta/\OmegaK$ and the corresponding source term is
\begin{equation}
	S_\text{relax} = -\rho\cv\frac{T-T_\text{init}}{t_\text{cool}} \Rightarrow \DP{T}{t} = -\frac{T-T_\text{init}}{\beta}\OmegaK.
\end{equation}
For simplicity, we adopt a constant $\beta=0.01$ in our models. This would correspond to a reference radius $R_0\sim50$--100\,au for typical disk parameters, placing our models in the radial range observable by ALMA \citep[e.g.,][]{andrews-etal-2018}, but our models are effectively compatible with locally isothermal models in the literature \citep[$\beta\rightarrow0$; e.g.,][]{stoll-etal-2017b,manger-klahr-2018,barraza-etal-2021}. Our choice of $\beta$ also results in a disk where the cooling criterion of \citet{lin-youdin-2015} is globally satisfied, and as a result the VSI operates at nearly full capacity \citep{manger-etal-2021}.

The disk is initialized in an equilibrium state according to Eqs.~\eqref{eq:power-law},~\eqref{eq:equilibrium1}~and~\eqref{eq:equilibrium2} with $p=-\nicefrac{3}{2}$, $q=-1$, and a constant $h=0.05$ for most models, with some models using $h=0.1$ instead. This corresponds to a nonflared disk with $T(R) = \{120.7,12.6\}$\,K at $\{5.2,50\}$\,au for a solar-mass star and a surface density profile $\Sigma(R)\propto R^{-\nicefrac{1}{2}}$. Our models are scale-free in terms of the reference radius $R_0$ and surface density $\Sigma_0=\Sigma(R=0)$, since we do not consider the back-reaction of gas onto the star. While our choice of parameters is not directly comparable to the temperature conditions at the ALMA range \citep[$q\approx-\nicefrac{1}{2}$, $h\approx0.1$, see for example][]{flock-etal-2020}, it allows us to construct models where the generated VSI turbulence corresponds to a constant effective $\alpha$ parameter \citep{shakura-sunyaev-1973} on the order of a few $10^{-4}$ similar to \citet{flock-etal-2020}, while also maintaining compatibility with previous studies \citep[e.g.,][]{nelson-etal-2013,stoll-etal-2017b,manger-etal-2020}. This becomes especially important when addressing the observability of VSI signatures, for which we compare our findings to those of \citet{barraza-etal-2021}.

\subsection{Numerics}
\label{sec:numerics}

We use the numerical hydrodynamics code \texttt{PLUTO 4.3} \citep{mignone-etal-2007} in a spherical polar $\{r, \theta, \phi\}$ geometry. \texttt{PLUTO} employs a finite-volume approach with a Riemann solver to integrate Eqs.~\eqref{eq:hydro1}--\eqref{eq:hydro3} in time. Our models cover the full azimuthal range $\phi\in[0,2\pi)$, a radial range $r\in[0.4,2.5]\,R_0$ and a vertical extent $z\in[-4, 4]\,H$, for a total of 8 scale heights from end to end vertically. After conducting a resolution study, we opted for a grid with 16 cells per scale height in the vertical and radial directions and 5 scale heights in the azimuthal direction, which translates to $N_r\times N_\theta \times N_\phi$ = $600\times128\times600$ cells for $h=0.05$. 
We found that this cell count adequately resolves both the growth and saturation phases of the VSI, while also keeping simulation runtimes reasonably feasible given their computational cost. A resolution analysis on our simulated VSI growth rates is provided in Appendix~\ref{appdx:resolution-study}.

Whenever possible, we use the FARGO module introduced by \citet{masset-2000} and implemented into \texttt{PLUTO} by \citet{mignone-etal-2012}. This yields a necessary speedup factor of $\sim$10 by subtracting the Keplerian background flow before solving the Riemann problem across all cell interfaces in the domain. We choose the HLLc Riemann solver \citep{harten-1983} with the Van Leer flux limiter \citep{vanleer-1974} and second-order reconstruction and time advancement schemes (options \texttt{LINEAR} and \texttt{RK2} respectively), but note that more accurate schemes such as third-order-accurate methods or less diffusive Riemann solvers and flux limiters yielded functionally identical results compared to our fiducial 2D model (discussed in Sect.~\ref{sec:2d}).

We apply wave-damping boundary conditions in the radial domain $R\in[0.4,0.5]\cup[2.1,2.5]\,R_0$ following \citet{devalborro-etal-2006}, using a damping timescale of ten orbits at the corresponding boundary edge. This suppresses the reflection of planet-generated spiral wakes while also providing a mass reservoir for the accreting, VSI-active disk. To further prevent waves of any kind from reentering the domain, we utilize ``strict outflow'' boundary conditions at the radial and polar directions, meaning that we allow outflow but not inflow of material at the boundary edge. In the polar direction, we extrapolate the density and pressure into the ghost cells assuming vertical hydrostatic equilibrium.

In models with an embedded planet, the latter is treated as a point-like gravitational potential with mass $\Mp =3\times10^{-4}\,\Mstar$ ($100\,\Mearth$ for a solar-mass star), smoothed over a distance of half its Hill sphere following the polynomial prescription of \citet{klahr-kley-2006}. We neglect gas back-reaction on both the planet and star. While in the linear regime we only expect the planet to launch spiral arms \citep{ogilvie-lubow-2002}, planets that are massive enough can open one or more gaps in their vicinity \citep{rafikov-2002}. The transition from one regime to the other can be defined using the disk thermal mass \citep{zhu-etal-2015}
\begin{equation}
	\Mth = \left.\frac{\cs^3}{\G \OmegaK}\right|_{\Rp} \approx 1\,M_\text{J}~\left(\frac{h_\text{p}}{0.1}\right)^3 \frac{\Mstar}{\Msun},
\end{equation}
which, for our choices of aspect ratio $h=\{0.05,0.1\}$ evaluates to $\{0.125,1\}\,M_\text{J}$. As a result, with our planet mass choice we can probe both ends of the planet--disk interaction spectrum, from the quasi-linear ($h=0.1$, $\Mp=0.3\,\Mth$) to the nonlinear regime ($h=0.05$, $\Mp=2.4\,\Mth$).

\begin{figure}
	\includegraphics[width=\linewidth]{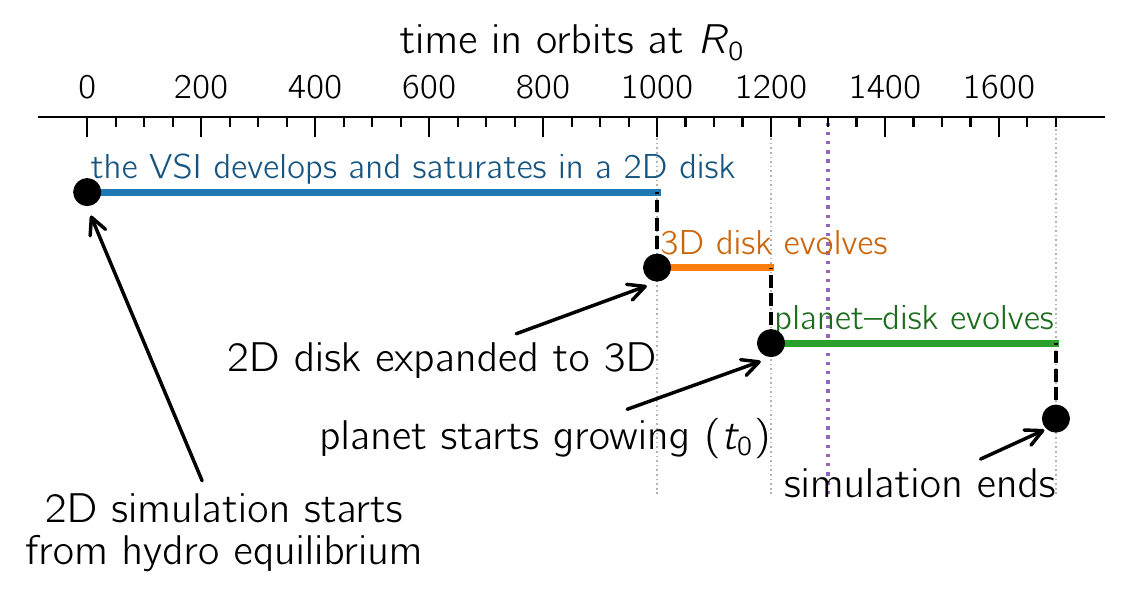}
	\caption{Timeline for our 3D models with embedded planets. Models in the blue, orange and green phases are discussed in Sects.~\ref{sec:2d}, \ref{sec:3d} and \ref{sec:3d-planet} respectively. The planet finishes growing by $t=t_0+100\,P_0$ (purple dotted line).}
	\label{fig:timeline}
\end{figure}

\subsection{Model timeline} %text done

Taking advantage of the axisymmetric nature of the VSI we first construct 2D models of the VSI on the $\{r, \theta\}$ plane. These models, while axisymmetric, involve all 3 velocity components. We then carry out a resolution analysis, in order to measure the turbulent viscosity $\alpha$ in a way compatible with previous 2D studies and verify that relevant VSI features in our 3D models are adequately resolved. While the VSI typically saturates within 30 local orbits \citep{nelson-etal-2013} which would translate to a required simulation time of $t_\text{2D}\approx2.5^{3/2}\,P_0$ $\approx$ 120 orbits at $R=R_0$, we continue until $t_\text{2D}=1000\,P_0$ to ensure the disk reaches a quasi-steady state while also allowing a reasonably long timeframe of 500 $P_0$ (between 500--1000\,$P_0$, with an output frequency of $1/P_0$) which we use to compute our time-averaged Reynolds stress once this quasi-steady state is reached. The 2D models are initialized at dynamical equilibrium through Eqs.~\eqref{eq:equilibrium1}~\&~\eqref{eq:equilibrium2}, with all velocity components perturbed by $1\%\,\cs$ to encourage the growth of VSI modes. We discuss our 2D results in Sect.~\ref{sec:2d}.

We then expand the disk into 3D by including the azimuthal direction. Similar to \citet{flock-etal-2020}, we duplicate the last state of our 2D models along the $\phi$-axis while also seeding the velocity with a further $1\%\,\cs$ perturbation to break the axisymmetry and allow the formation of azimuthal structure. The 3D models are then evolved for a further $200\,P_0$, and related results are presented in Sect.~\ref{sec:3d}.

Finally, a planet is inserted at $R=R_0$ in each of the now quasi-steady 3D disks and grows to its final mass over 100 orbits using the formula in \citet{devalborro-etal-2006}. The planet--disk models are then evolved for another $400~P_0$, until all relevant planet-generated features (spirals, gaps, vortices) have fully developed and the planet--VSI interplay can be assessed. Results from this phase of the timeline are discussed in Sect.~\ref{sec:3d-planet}.

A schematic of our model timeline is shown in Fig.~\ref{fig:timeline}. The 2D, 3D and planet--disk phases are color-coded in blue, orange and green, respectively.

\section{Results}
\label{sec:results}

In this section we present the results of our simulations. We begin by profiling the VSI, using an axisymmetric model to measure the growth rate and resulting stress levels in a saturated state. We then highlight the impact of 3D effects such as spiral arms and vortices on VSI activity. Finally, we analyze the effect of planet-generated features (spirals, vortices, rings and gaps) on VSI activity and attempt to disentangle the individual contributions of the planet and the VSI to the levels of turbulence we observe in our simulations.

\subsection{Reference model} %text done
\label{sec:2d}

\begin{figure*}[t]
	\includegraphics[width=\textwidth]{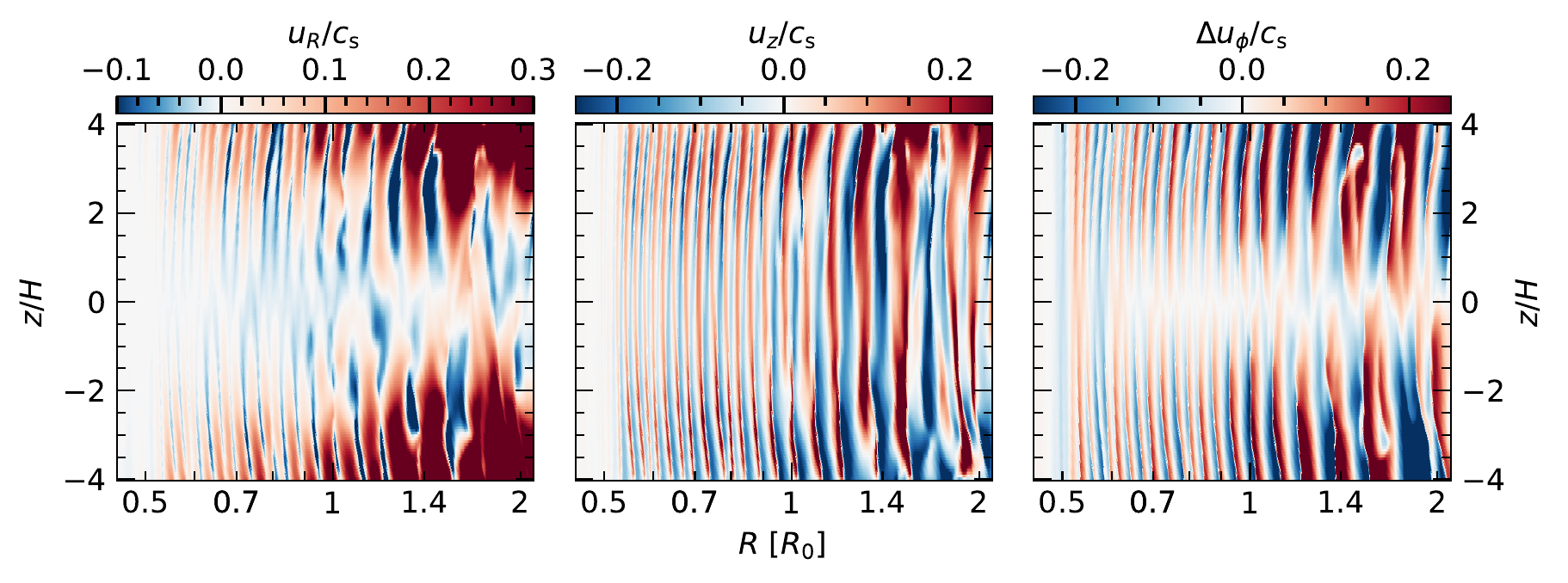}
	\caption{Radial ($u_R$), vertical ($u_z$), and azimuthal ($u_\phi$) gas velocity components in our fiducial axisymmetric model, highlighting the vertically elongated sheets of motion generated by the VSI. The Keplerian background $u_\text{K}$ has been subtracted from $u_\phi$, and all components have been normalized to the local sound speed.}
	\label{fig:vsi-example}
\end{figure*}

As has been shown before by numerous studies \citep[e.g.,][]{nelson-etal-2013,stoll-kley-2014,flock-etal-2017}, the VSI is characterized by quasi-axisymmetric, vertically elongated flows of gas that span the vertical extent of the disk while covering a proportionally much narrower distance in the radial direction. The result is the formation of channels of vertical gas motion between the two disk surfaces that turn over near each surface, before heading back toward the midplane. This is easily discernible through the vertical component of the gas velocity field $u_z$, which shows sheets of up- and downward motion along the $z$ direction throughout the disk (Fig.~\ref{fig:vsi-example}).

\begin{figure}[t]
	\includegraphics[width=\columnwidth]{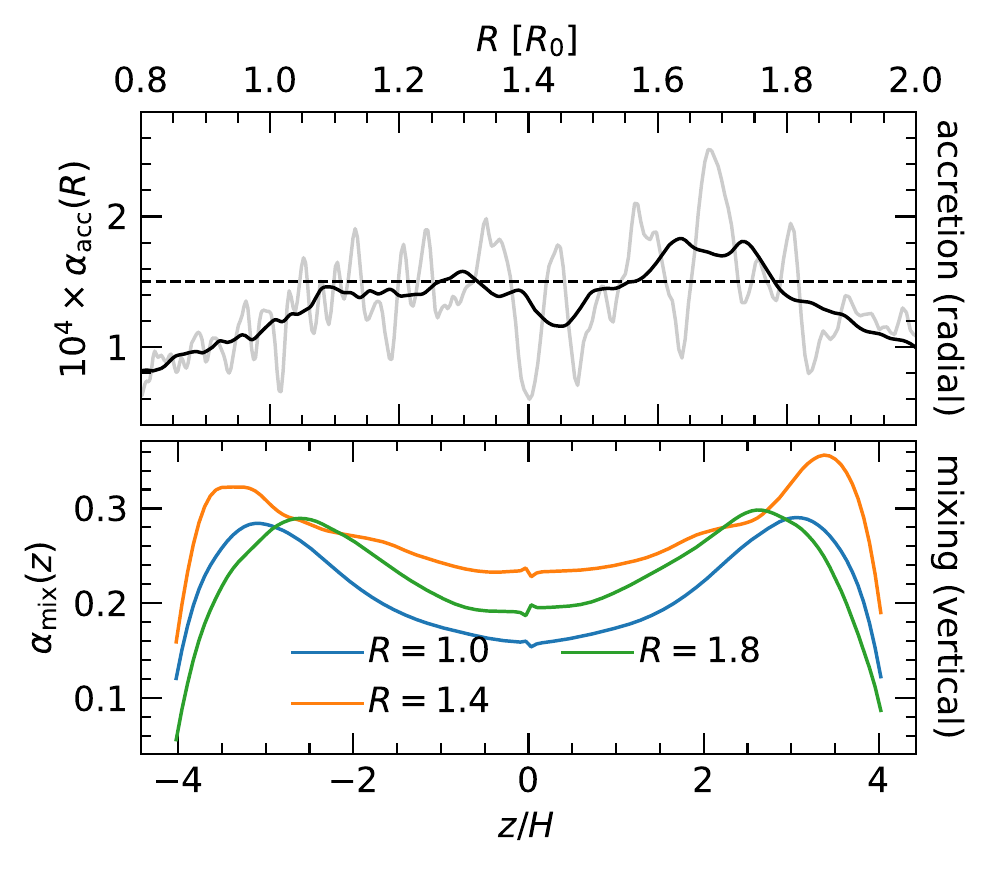}
	\caption{Radial and vertical profiles of the turbulent accretion parameter $\aacc$ and the turbulent mixing parameter $\amix$ in our fiducial axisymmetric model. We find a typical $\aacc\approx1.5\times10^{-4}$ and $\amix\approx 0.15\text{--}0.3$, in agreement with previous results. The faint curve on the upper panel refers to $\aacc$ without radial smoothing through Eq.~\eqref{eq:radial-smoothing}.}
	\label{fig:alpha-example}
\end{figure}

Consistent with previous studies, we compute a VSI-driven accretion stress $\aacc\approx1\text{--}3\times10^{-4}$ and a vertical turbulent mixing parameter $\amix\approx0.1\text{--}0.3$ for our choice of disk parameters. To measure $\alpha$ we adopt the method in \citet{balbus-papaloizou-1999}, which is similar to that of \citet{stoll-etal-2017b} albeit also accounting for the background velocity field (rather than assuming a stationary state with $\bar{u}_R=0$) and further adjusting the result by a factor of \nicefrac{2}{3}. Our method is described in detail in Appendix~\ref{appdx:tensors}, and our results for $\aacc$ and $\amix$ as functions of $R$ and $z$ respectively are shown in Fig.~\ref{fig:alpha-example}.

\FloatBarrier
\subsection{3D effects} %text done
\label{sec:3d}

While the VSI is primarily axisymmetric, including the azimuthal direction in fully 3D models can allow the growth of small-scale vortices at sufficient resolution \citep{richard-etal-2016,manger-klahr-2018} or trigger the Rossby-wave instability (RWI) at the radial location where the disk transitions from quiet to VSI-turbulent due to the formation of a gap edge \citep{flock-etal-2020}. Such vortices launch spiral arms as they decay \citep{rometsch-etal-2021}, generating a Reynolds stress that can contribute to the total accretion stress in the disk \citep{goodman-rafikov-2001}.

\begin{figure*}[t]
	\includegraphics[width=\textwidth]{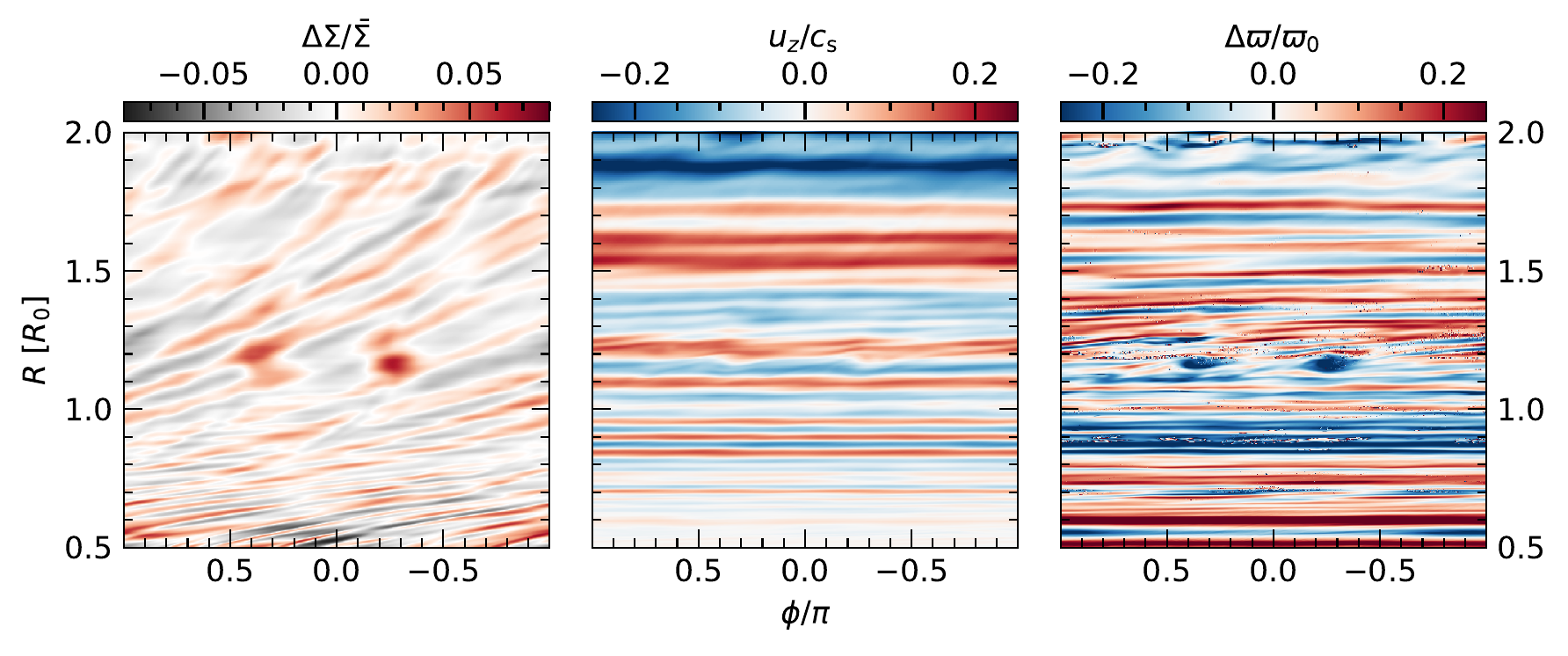}
	\caption{Highlights of our 3D fiducial model. Left: surface density deviations in the disk compared to the azimuthally averaged surface density $\bar{\Sigma}$, showing azimuthal structure in the form of spirals and vortices. Middle: vertical velocity of the gas at the disk midplane; quasi-axisymmetric stripes reveal VSI activity. Right: A map of the gas vortensity deviations compared to the initial (Keplerian) vortensity $\varpi_0$, highlighting two vortices orbiting at $R\approx1.15\,R_0$.}
	\label{fig:heatmap-3d}
\end{figure*}

Figure~\ref{fig:heatmap-3d} shows a snapshot of the gas surface density (left) and vertical velocity component (middle) for our fiducial 3D model that started from an axisymmetric, fully VSI-saturated state and has evolved for 200 orbits at $R=R_0$. The gas shows azimuthal structure, with spiral arms permeating the disk and two vortices orbiting at $R\approx1.15\,R_0$. Nevertheless, the disk is still VSI-active, as indicated by the largely axisymmetric velocity component $u_z$ with $u_z^\text{mid}\approx0.2\,\cs$, which is comparable to what we found in our axisymmetric VSI-saturated models in Fig.~\ref{fig:vsi-example}. To further highlight the presence of the vortices, we also plot the vortensity \citep{masset-llambay-2016}
\begin{equation}
	\label{eq:vortensity}
	\varpi = \left[\int\limits_{-\infty}^{+\infty}\frac{\rho}{(\nabla\times\vel)\cdot \hat{z}}\text{d}z\right]^{-1},
\end{equation}
on the right panel of the same figure.

In our 3D simulations we also observed the development of a parasitic instability in the inner boundary, likely caused by the damping of VSI modes on the interface between the active domain and the wave damping zone. This instability induced strong vertical motion near the interface of the inner damping zone ($R\approx0.4\,R_0$) and contributed to the formation of several small-scale vortices near the inner radial boundary. Similar to the vortices shown in Fig.~\ref{fig:heatmap-3d}, they launched spirals that spanned the entire domain, similar to what was observed by \citet{barraza-etal-2021}, likely contributing to the total Reynolds stress. We note that, in the planet--disk phase of our simulations, the planet overshadows the activity of such vortices. We provide more details on this phenomenon in Appendix~\ref{appdx:numerical-unhappiness}.

\begin{figure}[h]
	\includegraphics[width=\columnwidth]{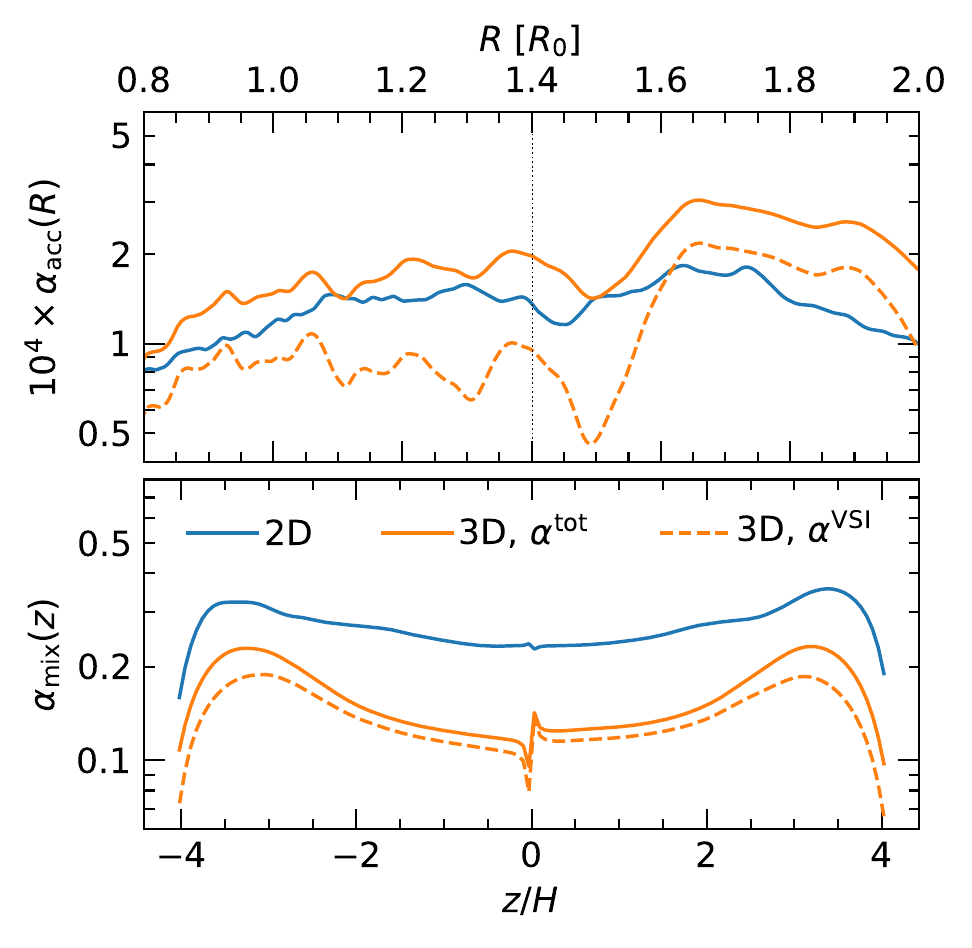}
	\caption{Turbulent parameters $\aacc$ and $\amix$ in our fiducial 3D model. Our 2D results from Fig.~\ref{fig:alpha-example} are shown in blue curves. The total stress in the disk is shown with solid curves, while dashed curves isolate the contribution of the VSI.
	While $\aacc^\tot$ increases slightly, $\amix$ decreases such that $\amix^\tot \approx \amix^\VSI$. This happens due to the stronger accretion stress supplied by spiral arms that in fact weaken the VSI, which is the primary driver of vertical motion.
	The bottom panel is calculated at the radial location shown by the vertical line in the top panel (here, $R=1.4\,R_0$).}
	\label{fig:stress-3d}
\end{figure}

We now measure turbulent stress by repeating the procedure in Sect.~\ref{sec:2d}, this time computing $\alpha$ using the full 3D gas data. As highlighted above and shown by the solid orange lines in Fig.~\ref{fig:stress-3d}, this method captures the accretion stress generated by the spiral arms seen in our 3D models and results in slightly higher $\aacc$ compared to our 2D models. However, the disk shows a weaker vertical mixing parameter $\amix$ in the region $R\sim R_0$, which is likely more strongly affected by the vortex present there, as well as in the region $R\lesssim0.7\,R_0$, which is dominated by the spirals launched by the vortices that formed near the inner boundary. 

By assuming that, approximately, the total stress is $\alpha^\text{tot} = \alpha^\text{VSI} + \alpha^\text{spiral}$, we attempt to isolate the contribution of the VSI to the total turbulent stress by recomputing $\alpha$ after first averaging our snapshots along the $\phi$ axis as explained in Appendix~\ref{appdx:tensors}. This results in spiral-free profiles, and the computed stress (dashed orange curves in Fig.~\ref{fig:stress-3d}) can be directly compared to the 2D models in Sect.~\ref{sec:2d} (blue curves) and contrasted against the fully 3D $\alpha$ stress (solid orange curves). We find that regions with a low $\amix^\text{tot}$ coincide with those with low $\amix^\text{VSI}$ and $\aacc^\text{VSI}$, even though $\aacc^\text{tot}$ stays consistently high due to spiral-driven Reynolds stress. In other words, while the spiral-rich 3D disk shows a slightly higher accretion stress \citep[as has been shown by previous 3D studies such as][]{manger-klahr-2018}, the contribution of the VSI to this total stress is in fact lower due to the spiral arms interfering with VSI modes, and vertical mixing in this 3D disk is lower than what an axisymmetric model would suggest. The latter supports the idea that the main driver of vertical mixing ($\delta u_z$) is the VSI, while the spiral arms launched by these relatively weak vortices mainly affect the radial and azimuthal components of gas velocity ($\delta u_R$, $\delta u_\phi$) instead, driving accretion.

Our argument, however, also implies that the spiral arms in our 3D runs should be responsible for part of the accretion stress observed. \citet{larson-1990} showed that spiral shocks (such as those present in our models) can produce an effective $\aacc^\text{spiral} \approx 0.013 h \sqrt{h+0.08}$, or $\approx2\times10^{-4}$ for $h=0.05$. This estimate is compatible with our results for $\aacc^\text{tot}$, and supports the idea that the observed accretion stress is driven by a combination of spiral arms and VSI modes in the disk.

\FloatBarrier

\subsection{Planet--VSI interaction}
\label{sec:3d-planet}

In the previous section we showed how spiral arms and vortex activity can inhibit the VSI by interfering with its vertical mixing capabilities while maintaining a high accretion stress. Here, we extend our models by embedding planets in our disks. While \citet{stoll-etal-2017a} have already examined the planet--VSI interplay as a function of planet mass, qualitatively showing that VSI activity is weaker near the planet for sufficiently massive planets, we intend to disentangle and quantify the contribution of the VSI to the total accretion and mixing Reynolds components compared to that of the planet and the features it generates (spirals, gaps, vortices).

Here, we focus on two models representative of the sub- and superthermal planet mass regimes by first showing the qualitative effect of the planet on the vertical velocity component of the disk, before calculating both $\alpha^\text{tot}$ and $\alpha^\text{VSI}$ in order to highlight the effect of the planet on the VSI. We then collect our estimates on $\aacc$ and $\amix$ with and without the contribution of nonaxisymmetric features for all 3D runs in Table~\ref{table:stress-all}.

\begin{table}[h]
	\caption{Estimates of turbulent $\alpha$ in our models. Planet mass is quoted in $\Mth$. Values for models with embedded planets serve as upper limits, especially for the last model where the VSI is only present in in the inner disk (see Fig.~\ref{fig:planet-stress-highmass}).}
	\label{table:stress-all}
	\centering
	\renewcommand{\arraystretch}{1.25} %for the \amix label
	\begin{tabular}{c | c | c | c | c | c}
		\hline
		$h$ & $\Mp$ & $\aacc^\text{tot}$ & $\amix^\text{tot}$ & $\aacc^\text{VSI}$ & $\amix^\text{VSI}$\\
		\hline
		\hline
		0.10 & --- & $4.51\times10^{-4}$ & $0.115$ & $2.53\times10^{-4}$ & $0.103$\\
		0.10 & 0.3 & $8.39\times10^{-4}$ & $0.058$ & $1.40\times10^{-4}$ & $0.045$\\
		0.05 & --- & $1.82\times10^{-4}$ & $0.128$ & $1.06\times10^{-4}$ & $0.117$\\
		0.05 & 2.4 & $1.32\times10^{-3}$ & $0.056$ & $4.21\times10^{-5}$ & $0.039$\\
	\end{tabular}
\end{table}

\subsubsection{Quasi-linear regime: $h=0.1\rightarrow\Mp=0.3\,\Mth$}
\label{sec:low-mass}

\begin{figure}[h]
	\includegraphics[width=\columnwidth]{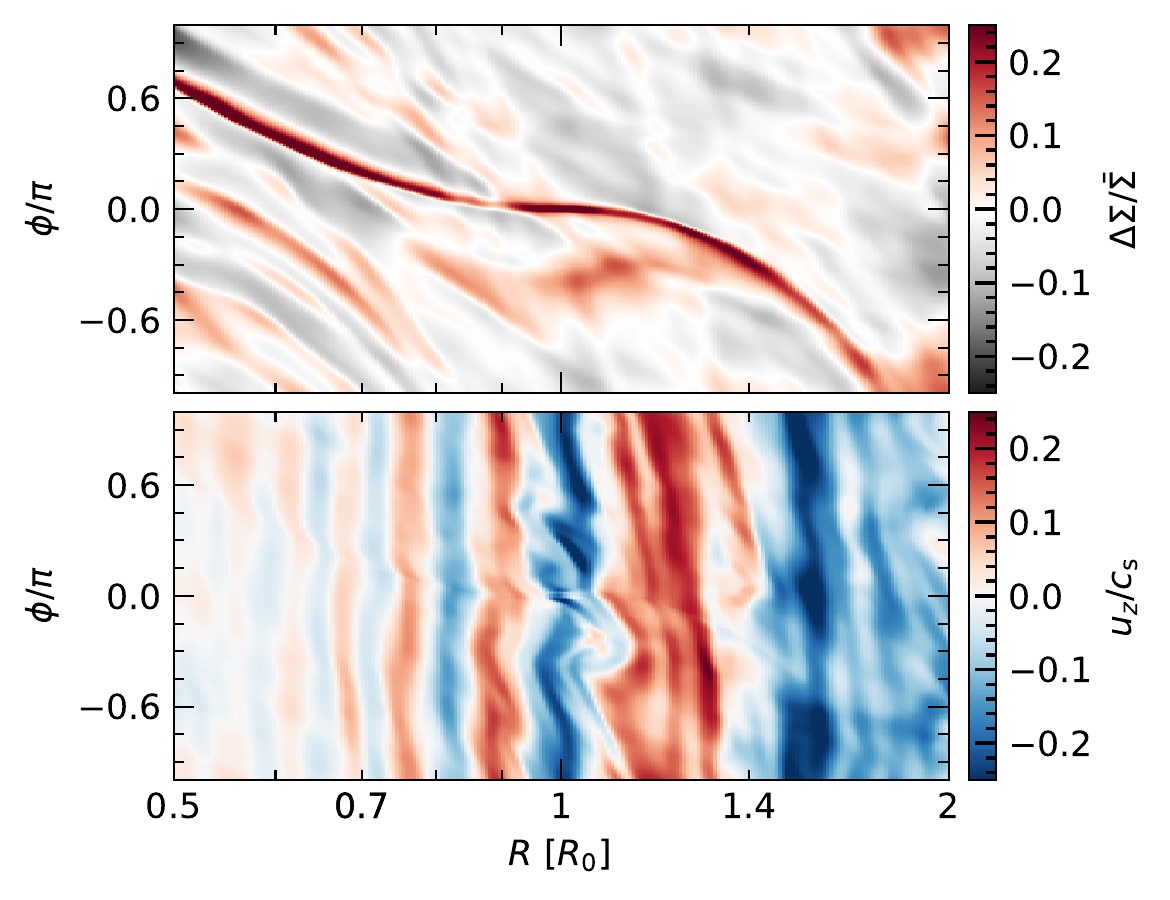}
	\caption{Surface density deviations and vertical velocity maps at the midplane for our disk with an embedded subthermal-mass planet ($h=0.1$, $\Mp=0.3\,\Mth$) at $R=R_0$ after 500 planetary orbits. While the planet's primary spiral is prominent in this color range, the VSI continues to operate as shown by the quasi-axisymmetric stripes of vertical motion in the bottom panel.}
	\label{fig:planet-state-lowmass}
\end{figure}

Figure~\ref{fig:planet-state-lowmass} shows an example of the typical disk structure around a subthermal-mass planet with $\Mp = 0.3\,\Mth$, taken after 100 planetary orbits in our model with $h=0.1$, $\Mp=100\,\Mearth$. While the planet's spiral arms are visible (top panel), the disk is otherwise weakly affected by the planet and the VSI remains operational, seemingly undeterred at least qualitatively (bottom panel).

Similar to our analysis in Sect.~\ref{sec:3d}, we now calculate the accretion stress $\aacc$ and vertical mixing parameter $\amix$ by first considering the full 3D structure of the disk (which includes the effects of spiral arms and results in a total stress $\alpha^\text{tot}$), and then comparing our findings to an estimate of the VSI-driven stress $\alpha^\text{VSI}$ by first averaging all quantities along the $\phi$ axis.

\begin{figure}[t]
	\includegraphics[width=\columnwidth]{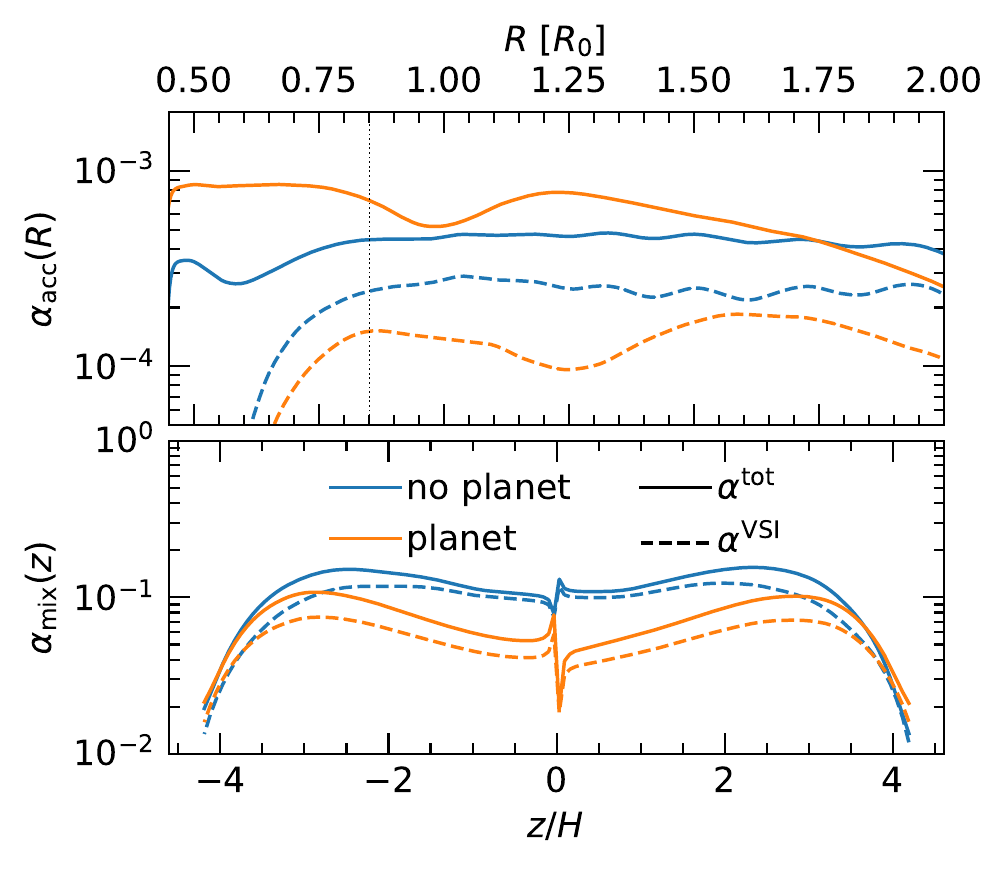}
	\caption{Turbulent parameters $\aacc$ and $\amix$ presented similar to Fig.~\ref{fig:stress-3d}, where the blue lines instead refer to the corresponding 3D model before the planet is introduced. As with Fig.~\ref{fig:stress-3d}, $\aacc$ increases but the overall mixing driven by the VSI is weaker by a factor of 2.}
	\label{fig:planet-stress-lowmass}
\end{figure}

Our results, shown in Fig.~\ref{fig:planet-stress-lowmass}, show that the presence of the planet does indeed affect both the total Reynolds stress in the disk, as well as the isolated contribution of the VSI to said stress. The total accretion stress $\aacc^\text{tot}$ (top panel, solid orange curve) increases when compared to the 3D run without an embedded planet (solid blue curve, see also the comparison between 2D and 3D for $h=0.05$ in Fig.~\ref{fig:stress-3d}) due to the spiral arms launched by said planet, in line with our expectations from Sect.~\ref{sec:3d}. At the same time, spiral activity interferes with the vertical velocity field, weakening VSI-driven stress $\aacc^\text{VSI}$ (dashed orange curve) by a factor of $\approx2$ compared to its contribution in the 3D run without a planet (dashed blue curve).
The same effect can be observed for the vertical mixing efficiency through $\amix$ on the right panel of Fig.~\ref{fig:planet-stress-lowmass}, as the VSI is the main driver of vertical motion (and thus $\amix$).

While the embedded planet can also excite vertical motion via buoyancy waves \citep{zhu-etal-2015} we could not find decisive evidence of their existence, most likely due to the quite low $\beta=0.01$.
An additional contribution to $\aacc$, also suggested by \citet{stoll-etal-2017a}, would very likely exist due to the spiral wave instability \citep[SWI,][]{bae-etal-2016a,bae-etal-2016b} producing an accretion stress with an equivalent $\aacc\sim 10^{-4}$ for this low planet mass.

All in all, in the quasilinear, subthermal-mass regime, we conclude that the VSI can coexist with a planet, albeit in a slightly weaker state. A model with $h=0.05$ and $\Mp=13\,\Mearth$, yielding $0.3\,\Mth$, resulted in very similar behavior, further highlighting that $\Mth$ is the relevant quantity to gauge the planet's effect on the disk regardless of disk model. In the next section, we investigate a scenario with a superthermal-mass planet.

\FloatBarrier

\subsubsection{Nonlinear regime: $h=0.05\rightarrow\Mp=2.4\,\Mth$}
\label{sec:high-mass}

\begin{figure}[t]
	\includegraphics[width=\columnwidth]{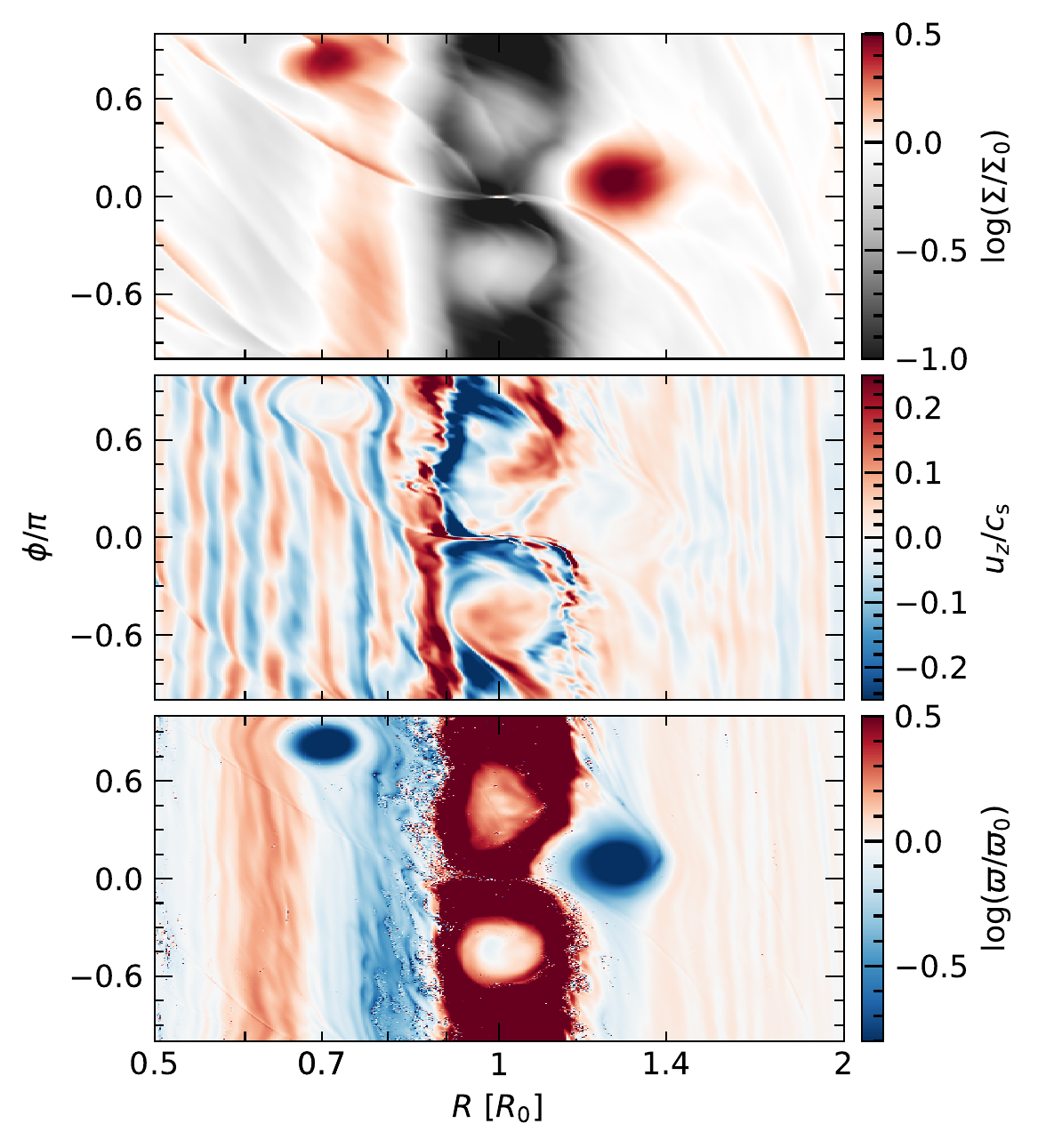}
	\caption{Two-dimensional heatmaps of relevant quantities similar to Fig.~\ref{fig:heatmap-3d}, showcasing the planet's effect on the disk after 500 orbits. The planet carves a deep gap and forms an additional, secondary ring (top) which lead to the formation of two massive vortices (top, bottom). VSI activity is significantly weaker in the outer disk but present in the inner disk (middle). The initial surface density and vortensity profiles are labeled with $\Sigma_0$ and $\varpi_0$.}
	\label{fig:planet-state-highmass}
\end{figure}

Contrary to the disk model with $h=0.1$ discussed in the previous paragraph, a planet with $\Mp=100\,\Mearth$ amounts to $2.4\,\Mth$ in a disk with $h=0.05$. This superthermal-mass planet, shown in Fig.~\ref{fig:planet-state-highmass}, shows significantly more features and affects the disk more strongly and on a larger radial scale. Here, the disk shows much more pronounced spirals as well as two gaps, with the center of the secondary gap agreeing very well with the position predicted by \citet{zhang-etal-2018} for a low-viscosity protoplanetary disk ($\aacc\lesssim5\times10^{-5}$). The conditions for the RWI are satisfied at the outer edge of both gaps, giving rise to two large-scale vortices. In particular, the massive vortex orbiting the planet's primary gap encloses a gas mass of $3.3\,\Mp=1\,\Mjup$ and survives over the entire duration of the simulation, continuously launching spiral arms that are comparable in magnitude to those formed by the planet into the outer disk.

\begin{figure}[h]
	\includegraphics[width=\columnwidth]{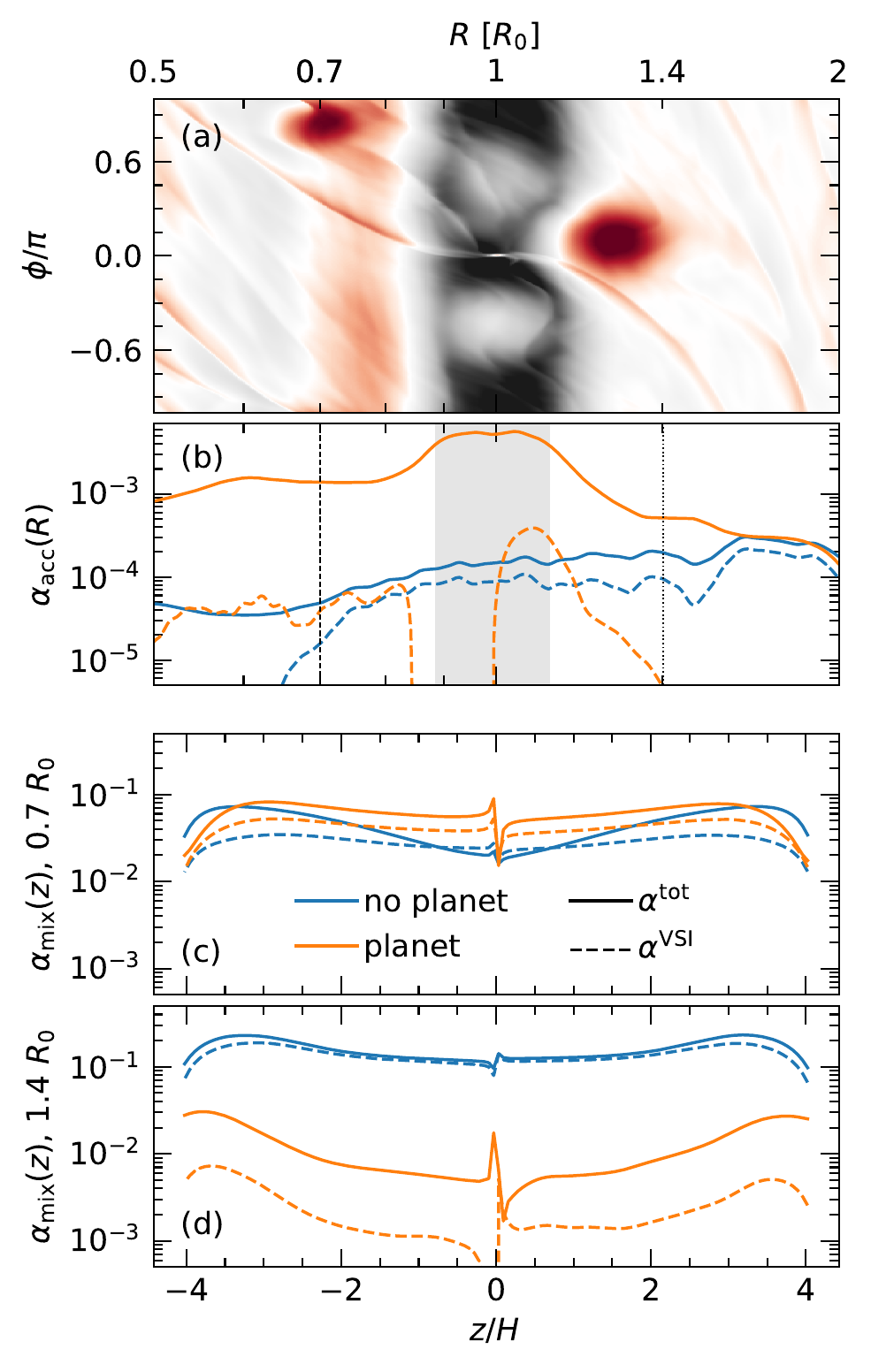}
	\caption{Turbulence parameters $\aacc$ and $\amix$ similar to Fig.~\ref{fig:planet-stress-lowmass}, but for the superthermal-mass case with $h=0.05$, $\Mp=2.4\,\Mth$. The surface density snapshot of Fig.~\ref{fig:planet-state-highmass} is also shown on panel \emph{a}, to highlight features present at different radii. The parameter $\aacc$ (panel \emph{b}) is dominated by the planet's spiral arms, but the VSI maintains a contribution in the inner disk (see also panel \emph{c}). In the outer disk, VSI activity is weaker by 1--3 orders of magnitude compared to its fiducial value (see also panel \emph{d}). A gray band marks the planet's gap region, defined as $R_\text{p}\pm 2.5\,R_\text{Hill}$.}
	\label{fig:planet-stress-highmass}
\end{figure}

In this regime, we see both qualitatively (Fig.~\ref{fig:planet-state-highmass}) and quantitatively (Fig.~\ref{fig:planet-stress-highmass}) that the VSI is significantly weaker, if not completely quenched, in the outer disk ($R\gtrsim R_0$, see in particular Fig.~\ref{fig:planet-state-highmass}--middle or Fig.~\ref{fig:planet-stress-highmass}--top). This is a result of the combined activity of the planet's strong spiral shocks near the planet, the massive vortex at the outer edge of the primary gap, and the presence of both planet- and vortex-driven spiral arms in the entire outer disk.

We further show that the VSI remains active in the inner disk ($R\lesssim R_0$), consistent with \citet{stoll-etal-2017a}, even though a vortex is present at the edge of the planet's secondary gap (see Fig.~\ref{fig:planet-stress-highmass}). This likely relates to the stronger vertical shear in the inner disk, which is proportional to the local $\OmegaK$, as well as the faster growth rate of the instability in the inner disk ($\propto\OmegaK^{-1}$) when compared to the timescale over which spiral arms, which corotate with the planet ($\propto\Omega_\text{p}^{-1}$), propagate radially through the gas. The quicker growth of the VSI in the inner disk, combined with the smaller, weaker vortex there compared to the massive vortex in the outer disk, allows the VSI to recover to levels comparable to those found in 2D or 3D runs without an embedded planet (see top and middle panels of Fig.~\ref{fig:planet-stress-highmass}).

Overall, a superthermal-mass planet in a nearly inviscid disk generates strong nonaxisymmetric features (spiral shocks, massive vortices and subsequent spirals) that can directly quench the VSI in their vicinity and especially so in the outer disk ($R\gtrsim\Rp$). We therefore conclude that the VSI is very unlikely to coexist with massive planets ($\Mp\gtrsim\Mth$), consistent with the qualitative findings of \citet{stoll-etal-2017a} and the synthetic observations by \citet{barraza-etal-2022}.

\FloatBarrier

\subsection{Vortex activity}
\label{sec:3d-vortex}

\begin{figure}[t]
	\includegraphics[width=\columnwidth]{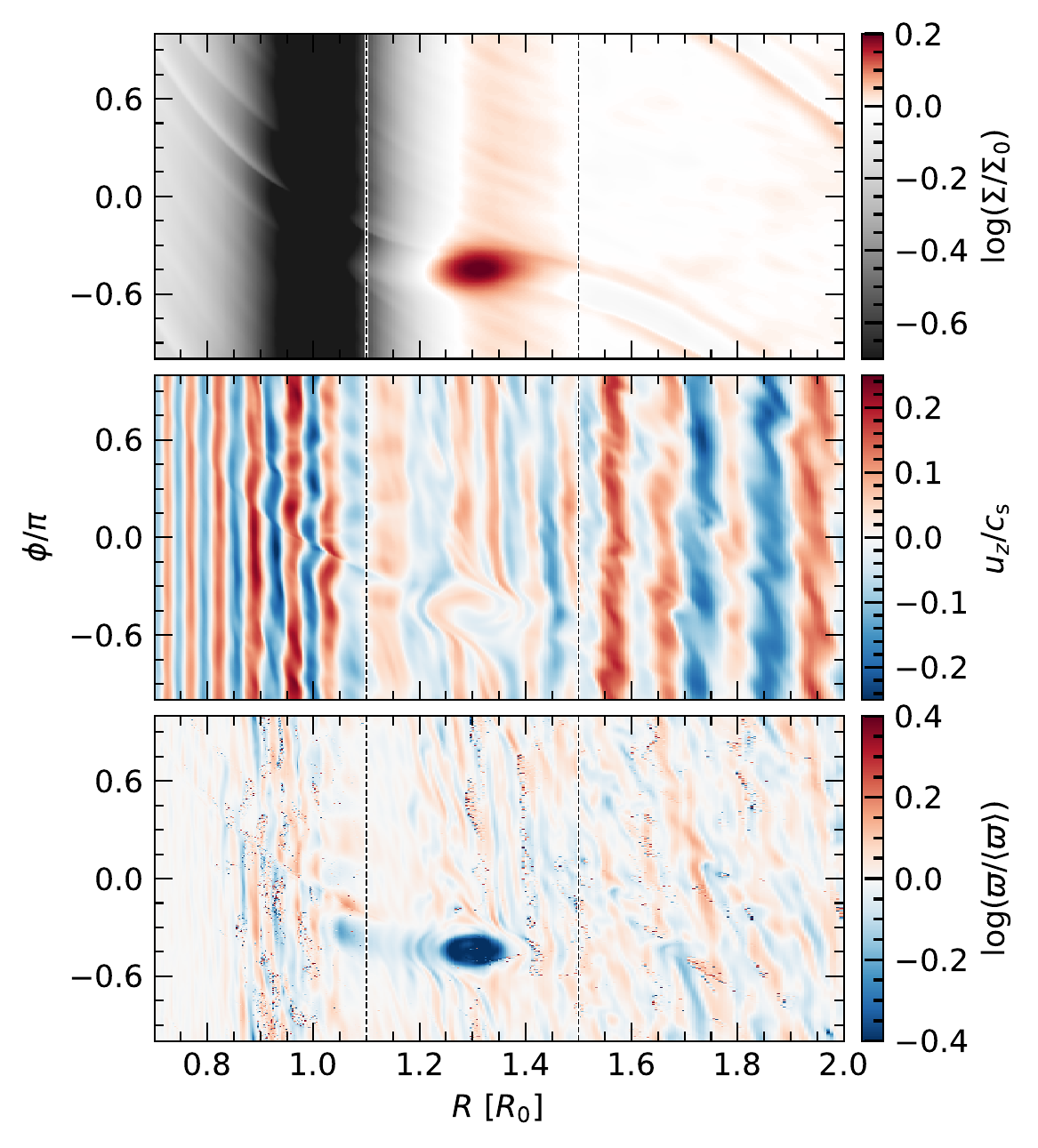}
	\caption{Surface density, vertical velocity and vortensity snapshots for our run with an embedded vortex after 400 orbits at $R_0$. To eliminate the stripes of vortensity deviations (see Fig.~\ref{fig:heatmap-3d}) and highlight the $\varpi$ minimum about the vortex, we normalize $\varpi$ to its azimuthal median $\mean{\varpi}$.}
	\label{fig:toy-vortex}
\end{figure}

To further elucidate the effect of vortices and their spiral arms that was discussed in the previous section, we construct an additional 3D model based on our fiducial setup ($h=0.05$), with the goal being to artificially spawn a massive vortex into the disk and measure the turbulent stress in its vicinity similar to Sects.~\ref{sec:3d}~and~\ref{sec:3d-planet}. We initialize the disk with a circular density gap centered at $R=R_0$, and place a blob of gas with total mass $1\,\Mjup$ at the outer gap edge using a Gaussian distribution centered at $R=1.25\,R_0$, $\phi=\pi$ (see top panel of Fig.~\ref{fig:toy-vortex})
\begin{equation}
	\rho(R,\phi) = \rho^\text{eq}\,\times\,f_\text{gap}(R)\,\times\,f_\text{vortex}(R, \phi),
\end{equation}
with $f_\text{gap}$ and $f_\text{vortex}$ motivated by our embedded-planet models in Sect.~\ref{sec:3d-planet}:
\begin{equation}
	\begin{split}
		f_\text{gap}(R) = 1 - 0.95 &\exp\left[-\frac{(R/R_0-1)^2}{0.05}\right]\\
		f_\text{vortex}(R,\phi) = 1 + 3.5&\exp\left[-0.5\left(\frac{R/R_0-1.25}{0.1}\right)^2\right] \\
		\times&\exp\left[-0.5\left(\frac{\phi-\pi}{0.35}\right)^2\right].\\ 
	\end{split}
\end{equation}

The gas mass is initialized with Keplerian vorticity and quickly shears into small gas clumps spread along the gap edge due to the radially-varying $\Omega(R)$ profile. At the same time the RWI, the conditions for which are satisfied at the gap edge, causes the formation of small-scale vortices at the location of each clump, which over time coalesce into a single vortex. The result is the formation of a massive vortex that orbits at the outer gap edge, launching spiral arms through the otherwise VSI-active disk.

\begin{figure}[h]
	\includegraphics[width=\columnwidth]{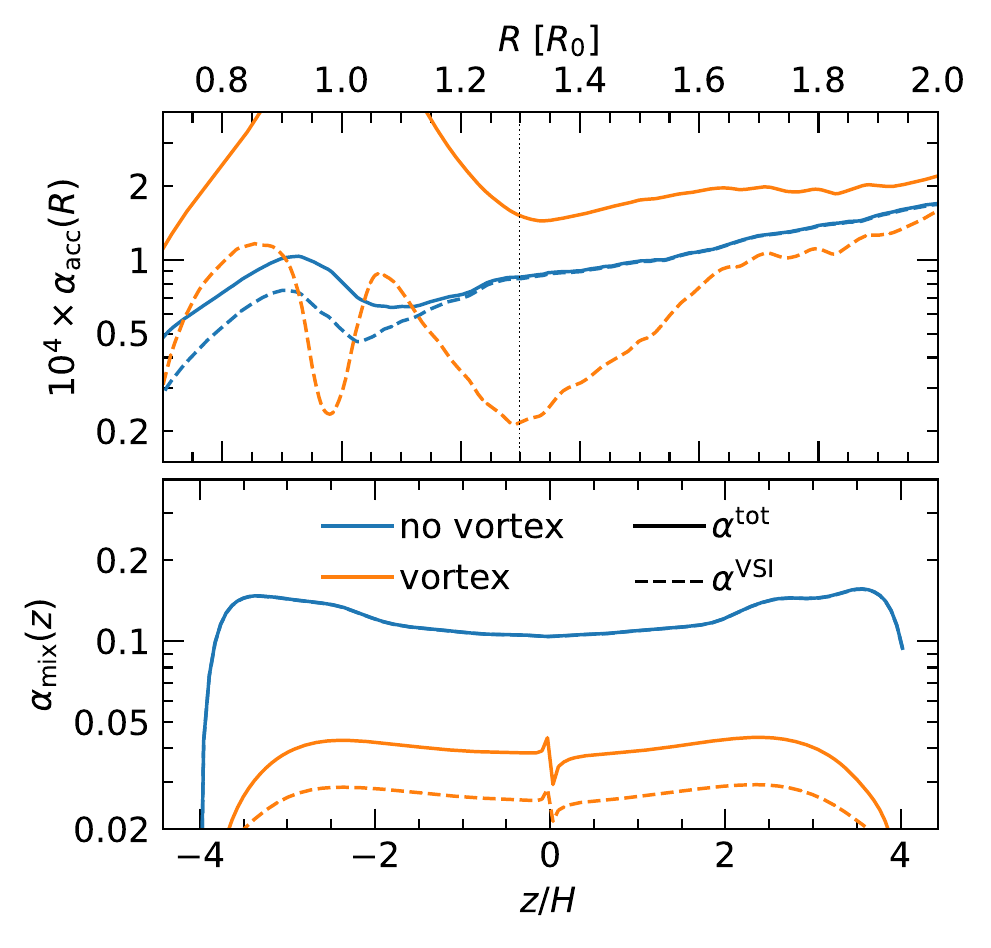}
	\caption{Turbulent $\aacc$ and $\amix$ in our vortex runs, similar to Fig.~\ref{fig:stress-3d}. VSI activity is significantly lower near the radial vicinity of the vortex ($R\approx 1.3\,R_0$).}
	\label{fig:toy-stress}
\end{figure}

Fig.~\ref{fig:toy-stress} shows the results of our data reduction for these models. Consistent with our findings in Sects.~\ref{sec:3d}~and~\ref{sec:3d-planet}, we find that $\aacc$ is slightly stronger in the model with a vortex orbiting the gap edge due to the spirals generated by the vortex. At the same time, $\amix$ drops near the vortex, signaling weaker VSI activity due to the disturbance of the gas velocity field caused by the vortex.We also highlight that the drop in gas density inside the gap does not seem to affect $\alpha$ in any noticeable way.

\FloatBarrier

\section{Discussion}
\label{sec:discussion}

In this section we justify our assumptions regarding our cooling prescription and discuss our findings with regard to observations of the VSI. We also briefly relate our findings to previous studies.

\subsection{Radiative effects}

Treatment of radiative effects---be it through a local, opacity-dependent parametrized cooling prescription \citep[e.g.,][]{pfeil-klahr-2021} or a full treatment of radiation transport via flux-limited diffusion \citep[FLD,][]{stoll-kley-2014,flock-etal-2020}---will certainly influence the planet's gap-opening capabilities \citep{ziampras-etal-2020b,miranda-rafikov-2020b}. Nevertheless, the position of spirals will remain largely unaffected and the resulting changes in the overall temperature profile will be minuscule in the regime where the VSI is expected be active \citep[$R\gtrsim10$\,au,][]{flock-etal-2020} due to the significantly weaker spiral shock heating at these distances from the star \citep{rafikov-2016,ziampras-etal-2020a}. For this reason, and since our topic of interest is the planet--VSI interaction in a fully VSI-active disk where radiative effects have been shown to not affect VSI activity significantly \citep{stoll-kley-2014,manger-etal-2021}, we do not carry out FLD models in 3D.

\subsection{On the observability of the VSI}

\citet{barraza-etal-2022} showed that the observational signature of the VSI in a disk with an embedded massive planet is significantly weaker by simulating ALMA gas kinematics observations. \citet{stoll-etal-2017a} similarly showed that, using $u_\text{z}$ as a tracer for VSI activity, the instability should be weaker in the outer disk for $\Mp=100\,\Mearth$, or $2.4\,\Mth$ in their models, but did not measure the vertical mixing parameter $\amix$ in their simulations and instead focused on accretion stress ($\aacc$).

Our models, for which we provide quantitative values of $\amix$, support the findings of both studies and enrich them further by providing insight into the effect of a large-scale vortex on $\amix$. This agreement also supports the robustness of our method of isolating the contribution of the VSI to the total stress by simply azimuthally averaging our data before computing the turbulent stress. It is also worth noting that our vertical profiles of $\amix(z)$ are approximately constant, which means that the VSI does not recover at large heights (see for example Fig.~\ref{fig:toy-stress}) even though features such as vortices are significantly denser near the midplane, with no sign of their activity near the disk surfaces.

This raises, however, the question whether VSI signatures could be observable in protoplanetary disks with azimuthal structures such as spiral arms, vortices, or planets. The findings of \citet{barraza-etal-2021} suggest that the vertical motion induced by the VSI could be observable in the near future in disks where the VSI can operate unhindered and determine the velocity field of the gas. By combining our results with those of \citet{stoll-etal-2017a} and \citet{barraza-etal-2022}, we conclude that it might be possible, albeit challenging, to observe VSI activity in disks with subthermal-mass planets or vortices and for typical disk parameters for ALMA ($h\sim0.1$, $q\approx-0.5$, such that $\alpha^\text{VSI}\sim10^{-4}$) in the future, should the instability be operating in the disk. However, we also deduce that it is unlikely that VSI signatures could be detected in the exterior regions of a disk containing a planet whose mass exceeds the thermal mass, since the planet-induced strong spiral shocks quench the activity of the instability.

\subsection{On the lifetime of vortices in our models}

In our model with $\Mp =2.4\,\Mth$ a planet opened two gaps, the outer edge of which grows RWI-unstable and spawns large-scale vortices. In particular, the vortices that form outside of the planet's primary gap grow to masses of up to $3.3\,\Mp=1\,\Mjup$. Given the nearly-locally isothermal disk conditions ($\beta=0.01$) and the lack of VSI activity in their vicinity resulting in a nearly inviscid disk with $\alpha \sim \max(\alpha_\text{num}, 10^{-7}) \ll 10^{-4}$, such vortices are expected to survive for tens of thousands of planetary orbits \citep{rometsch-etal-2021}.

However, the vortex growing on the outer edge of the secondary gap is smaller in comparison. This vortex, also visible in Fig.~\ref{fig:planet-state-highmass}, is not massive enough to strongly affect the vertical velocity field in its vicinity (bottom panel of Fig.~\ref{fig:planet-state-highmass}), and as a result it is effectively embedded in a VSI-active region with $\alpha\approx5\times10^{-5}$ (see top panel of Fig.~\ref{fig:planet-stress-highmass}), where it decays over $\approx10^3$ local orbits at $R\approx0.7\,R_0$. This lifetime is consistent with the estimates of \citet{rometsch-etal-2021} for a disk with $\alpha=10^{-4}$, $\beta=0.01$ and $h=0.05$, suggesting that the turbulence driven by the VSI is compatible with 2D laminar-$\alpha$ models. This last point complements the findings of \citet{stoll-etal-2017a}, who showed that $\alpha$- and VSI-driven disks behave very similarly in terms of planet gap opening and torques exerted on embedded planets.

\subsection{On resolution and grid requirements for planet--VSI modeling}

Recent work by \citet{flores-etal-2020} has shown that the VSI is subject to the Kelvin--Helmholtz instability on the interface between its vertically-shearing corrugation modes, and \citet{cui-latter-2022} have further explored the nonlinear saturation of the VSI via a parametric instability. Both studies highlight the necessity of high-resolution modeling ($\gtrsim200$--300 cells per scale height) in order to fully resolve the interplay between all mechanisms present in VSI-active disks. Since we are mainly interested in the stress generated by the VSI and the planet--VSI interaction, which can be resolved adequately with our grid resolution, and for computational reasons, we chose not to run models at a higher resolution.

In addition, \citet{svanberg-etal-2022} have shown that the VSI saturates into coherent, large-scale, outwardly-traveling inertial waves. In this framework, the planet can be seen as a potential that disrupts this outward propagation of VSI waves beyond its semimajor axis. While our data supports this idea, with VSI stress plummeting in the outer disk in our models with $\Mp=\Mth$ (Sect.~\ref{sec:high-mass}), it is difficult to explore this scenario without actually carrying out simulations with a large enough radial range that can accommodate the several radial zones that the VSI partitions the disk into.

Based on the above, a much higher grid resolution as well as larger radial domain would be necessary to resolve the global structure of the VSI and the instabilities that can attack it. The interplay between the VSI and the mechanisms covered in the above studies in disks with embedded planets is worth exploring in the future.

\section{Conclusions}
\label{sec:conclusions}

%what we did
We executed 3D numerical hydrodynamics simulations to study the vertical shear instability (VSI) and its interaction with nonaxisymmetric features such as embedded planets or vortices in a protoplanetary disk. We then isolated the contribution of the VSI to the total radial accretion and vertical mixing turbulence parameters ($\aacc$ and $\amix$ respectively) and quantified to what extent a feature such as a planet or vortex can interfere with VSI activity.

%how we did it
To achieve this, we first computed the total accretion and mixing parameters using the full 3D structure of a snapshot ($\alpha^\text{tot}$), which includes the contribution of nonaxisymmetric features. We then compared this to the stress calculated using azimuthally averaged datasets ($\alpha^\text{VSI}$), taking advantage of the inherently axisymmetric nature of the VSI.

%compare 2D to 3D
By first comparing a typical 2D axisymmetric VSI model to a fully 3D one, we found that the total accretion stress in 3D is slightly higher than in the 2D case, consistent with previous studies \citep{manger-klahr-2018}. We attribute this additional stress to the spiral arms that form in the 3D model, and highlight that the VSI-induced stress is in fact lower, as can be seen in both $\aacc$ and $\amix$. This is not surprising, as spiral arms can contribute strongly to accretion \citep{goodman-rafikov-2001} while the VSI remains the main driver of vertical motion in the disk. As a result, we find that $\aacc^\text{VSI} < \aacc^\text{2D} < \aacc^\text{tot}$ and $\amix^\text{VSI} \approx \amix^\text{tot} < \amix^\text{2D}$.

%planet case
In models with embedded planets, the spiral wakes excited by the planet can strongly affect the velocity field in the disk, generating an effective accretion stress. The result, similar to the 3D case without a planet, is slightly weaker vertical mixing while the radial accretion stress becomes higher, in line with the expectations from theory \citep{goodman-rafikov-2001}.

In addition, due to the nearly inviscid disk conditions, a planet with $\Mp\gtrsim\Mth$ will open a gap in the gas \citep{crida-etal-2006} while also generating strong zonal flows near its vicinity \citep{bi-etal-2021}, as well as large-scale vortices at its gap edge \citep{paardekooper-etal-2010,rometsch-etal-2021} that launch waves of similar magnitude as they decay over time. All of these processes can, to different extents, disrupt the velocity field in such a way that the contribution of the VSI to the total accretion stress is significantly weaker in the outer disk for a massive planet. In the inner disk, due to the absence of massive vortices, the stronger vertical shear rate, and the shorter growth timescale of the VSI ($\propto \OmegaK^{-1}$) in contrast to the ``disturbance timescale'' by the planet's spirals as they propagate radially inward ($\propto\Omega_\text{p}^{-1}$), the VSI remains active at levels similar to those expected by axisymmetric models, without embedded planets.

Additional mechanisms that a planet excites and can drive vertical motion, such as the spiral wave instability \citep[SWI,][]{bae-etal-2016a,bae-etal-2016b} or buoyancy waves \citep{zhu-etal-2015,mcnally-etal-2020}, could further hinder the contribution of the VSI to the total accretion and mixing turbulent parameters. In our case, the short cooling timescale of $\beta=0.01$ that we impose should minimize the excitation of buoyancy waves, while our grid resolution might be too low to reliably invoke the SWI as an alternative driver of turbulence. A study on the effect of either mechanism on the activity of the VSI is unfortunately out of the scope of our project, but is worth investigating in the future.

%vortex case
Vortices can, in principle, form without the necessity of the presence of a planet in VSI-active disks \citep{richard-etal-2016,manger-klahr-2018,flock-etal-2020}. A large-scale vortex can, similar to a planet, launch spiral arms that corotate with its orbit. Such spirals will also drive an accretion stress with $\aacc\sim10^{-4}$ for our disk parameters \citep{larson-1990}, and can subject the gas to vertical motion through one or more of the mechanisms outlined above. This results in a weaker VSI stress at the immediate vicinity of the vortex, that recovers at sufficient distances ($\approx4H$ from the eye of the vortex for our models). While this would very likely render the VSI undetectable and the mm dust layer considerably flatter at the vicinity of the vortex, the effect is weaker than in the case of an equally massive planet.

%closing remarks
In summary, our findings suggest that the VSI will be severely hindered in disks with massive embedded planets, and considerably weaker in the vicinity of large-scale vortices. This has implications for the height of the mm dust layer \citep{dullemond-etal-2022}, which is commonly the target of observations using the ALMA array \citep[e.g.,][]{andrews-etal-2018}, as well as for the observability of the instability using ALMA detections of molecular line emission \citep{barraza-etal-2021,barraza-etal-2022}.

% \newpage
\begin{acknowledgements}
AZ and RPN are grateful for the guidance and friendship of Willy Kley, and dedicate this paper to his memory. We thank the anonymous referee for their helpful suggestions and comments.
The authors acknowledge support by the High Performance and
Cloud Computing Group at the Zentrum f\"ur Datenverarbeitung of the University of T\"ubingen, and the state of Baden-W\"urttemberg through bwHPC. This research utilized Queen Mary's Apocrita HPC facility, supported by QMUL Research-IT (http://doi.org/10.5281/zenodo.438045). This work was performed using the DiRAC Data Intensive service at Leicester, operated by the University of Leicester IT Services, which forms part of the STFC DiRAC HPC Facility (www.dirac.ac.uk). The equipment was funded by BEIS capital funding via STFC capital grants ST/K000373/1 and ST/R002363/1 and STFC DiRAC Operations grant ST/R001014/1. DiRAC is part of the National e-Infrastructure.
AZ and RPN are supported by STFC grant ST/P000592/1, and RPN is supported by the Leverhulme Trust through grant RPG-2018-418.
All plots in this paper were made with the Python library \texttt{matplotlib} \citep{hunter-2007}.
\end{acknowledgements}

\newpage
\bibliographystyle{aa}
\bibliography{refs}

\begin{appendix}

\section{Measuring turbulent stress}
\label{appdx:tensors}
	
To measure turbulent stress we followed a four-step process based on \citet{balbus-papaloizou-1999}. In the section below, we denote a quantity $q$ averaged with respect to $y$ as $\mean{q}_y$.
	
We begin by first computing an axisymmetric, time-averaged disk state for all quantities $\rho$, $\vel$ and $P$ using 150--500 snapshots depending on model, taken at intervals of $1/P_0$. For a quantity $q$, we have
\begin{equation}
	\mean{q}_{\phi t} = \frac{1}{N_\text{files}}\sum\limits_{i=i_\text{start}}^{i_\text{end}}\mean{q^i}_\phi,\quad\mean{q^i}_\phi=\frac{1}{2\pi}\int\limits_{0}^{2\pi}q^i\,\text{d}\phi,
\end{equation}
where $i_\text{start}$ and $i_\text{end}$ denote the indices of the first and last snapshots used, and $N_\text{files} = i_\text{end}-i_\text{start}+1$ is the number of snapshots.

We then compute the $R\phi$ and $z\phi$ components of the Reynolds stress tensor $\tensor{T}(\bm{r})$
\begin{equation}
	\label{eq:stress-components}
	T_{x\phi}^i(\bm{r}) = \rho^i\cdot\left[u_x^i -\mean{u_x}_{\phi t}\right]\cdot\left[u_\phi^i-\mean{u_\phi}_{\phi t}\right],\qquad x\in\{R, z\},
\end{equation}
and obtain two 3D data cubes $\mean{T_{R\phi}}_t$ and $\mean{T_{z\phi}}_t$ by averaging over the same snapshots that we used to compute the mean, background field. We then compute the azimuthal average and apply a radial smoothing over a window of a local scale height $H(R)$ on both datasets:
\begin{equation}
	\label{eq:radial-smoothing}
	T^\prime_{x\phi}(R, z) = \frac{1}{\Delta R}\int\limits_{R-\Delta R/2}^{R+\Delta R/2}\mean{T_{x\phi}}_{t\phi}\,\text{d}R^\prime,\qquad\Delta R = H(R).
\end{equation}
This is equivalent to a radial rolling average, and can be easily implemented with a first-degree Savitzky--Golay filter (available as \texttt{scipy.signal.savgol\_filter} in \texttt{Python}).

To compute $\aacc(R)$ we integrate $T^\prime_{R\phi}$ vertically and finally obtain
\begin{equation}
	\aacc(R) = -W_{R\phi} \cdot \left(R\D{\Omega}{R}\Sigma \cs H\right)^{-1},\qquad W_{R\phi}=\int\limits_{z_\text{min}}^{z_\text{max}} T^\prime_{R\phi}\,\text{d}z,
\end{equation}
where $\Sigma$ is computed using the time- and azimuthally-averaged $\mean{\rho}_{t\phi}$. We note that, since the mean azimuthal velocity and disk temperature do not change significantly due to the VSI ($\Delta u_\phi/{u_\phi}_0\sim 10^{-2}$, $\Delta T/T_0\sim 10^{-6}$), one can also instead use the initial, analytical profiles of $\Omega$, $\cs$ and $H$ in these calculations. However, since the disk is constantly accreting (i.e., losing mass in our models), the mean density $\mean{\rho}_{t\phi}$ is more appropriate than the initial profile $\rho^\text{eq}$ (see Eq.~\eqref{eq:equilibrium1}) when calculating $\Sigma$.

Finally, we compute $\amix$ by replacing the viscous stress tensor component in Eq.~(10) of \citet{stoll-etal-2017b} with $T_{z\phi}^\prime$ and find
\begin{equation}
	\amix = \frac{T_{z\phi}^\prime}{\mean{P}_{\phi t}} \cdot\left(\frac{|q| h}{2}\frac{z}{H}\right)^{-1}.
\end{equation}
We note that this only provides an estimate of $\amix$, as it is not backed by the accretion model by \citet{balbus-papaloizou-1999}. A more appropriate method of calculating $\amix$ would involve scattering dust particles in a numerical model and measuring their scale height as they are lifted by the gas \citep[e.g.,][]{dullemond-etal-2022}. Nevertheless, our method provides compatible results with those of \citet{dullemond-etal-2022} for our models.

To isolate the contribution of the VSI, we simply replace $\rho^i$ and $u^i$ in Eq.~\eqref{eq:stress-components} with their azimuthal averages $\mean{\rho^i}_\phi$ and $\mean{u^i}_\phi$ respectively. This means that azimuthal averaging in Eq.~\eqref{eq:radial-smoothing} is not necessary, as $\mean{T_{x\phi}}_{t}$ is already axisymmetric in this case.

\section{Resolution study}
\label{appdx:resolution-study}

\begin{figure}[h]
	\includegraphics[width=\columnwidth]{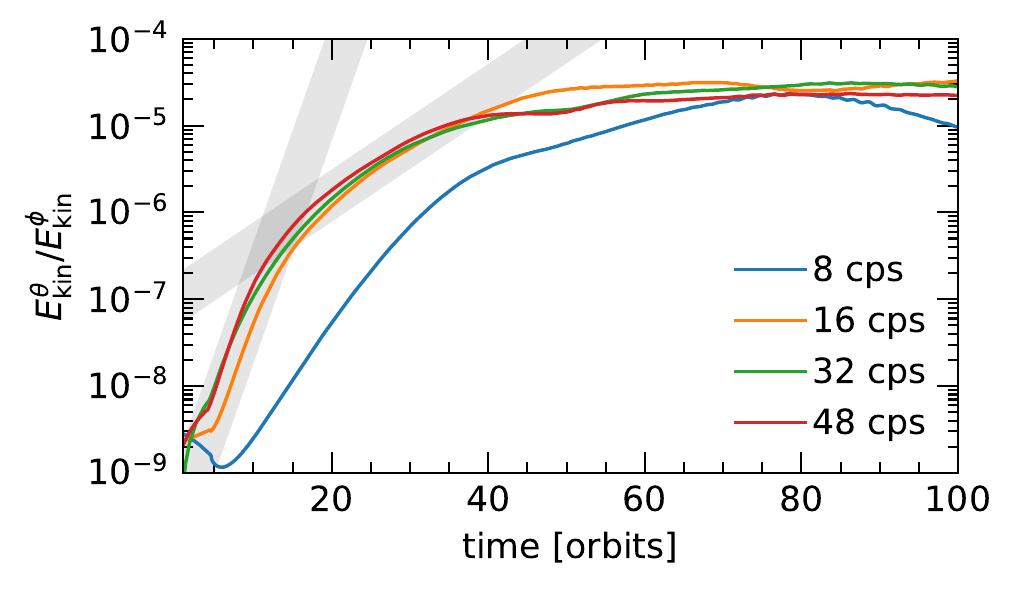}
	\caption{Growth rates of the VSI as a function of resolution in cells per scale height (cps) in our axisymmetric models. The two shaded bands denote a growth rate of 0.58/orbit and 0.13/orbit for the steep and shallow bands respectively, in agreement with \citet{stoll-kley-2014}.}
	\label{fig:res-growth-rates}
\end{figure}

Previous studies have shown that a resolution of 16 cells per scale height (cps) is enough to resolve the growth and saturation of the VSI and recover a value of $\alpha\sim 10^{-4}$. Nevertheless, to verify that the VSI is resolved appropriately in our models, we conduct a short resolution study where we measure the growth rate (Fig.~\ref{fig:res-growth-rates}) of the instability using 2D axisymmetric models.

After a local linear stability analysis we found that the fastest-growing VSI mode has a wavelength of approximately $0.1\,H$, which requires a radial resolution of 20 cps in order to be resolved over 2 cells. To maintain compatibility with previous studies, we then run models at 8, 16, 32 and 48 cps in both $r$ and $\theta$ directions.

We find that a resolution of 16 cps captures a growth rate of $\approx0.58/\text{orbit}$, similar to that reported by \citet{stoll-kley-2014}, with higher-resolution models showing very similar results (see Fig.~\ref{fig:res-growth-rates}).

\section{Issues with the inner boundary}
\label{appdx:numerical-unhappiness}

\begin{figure}[h]
	\centering
	\includegraphics[width=\textwidth]{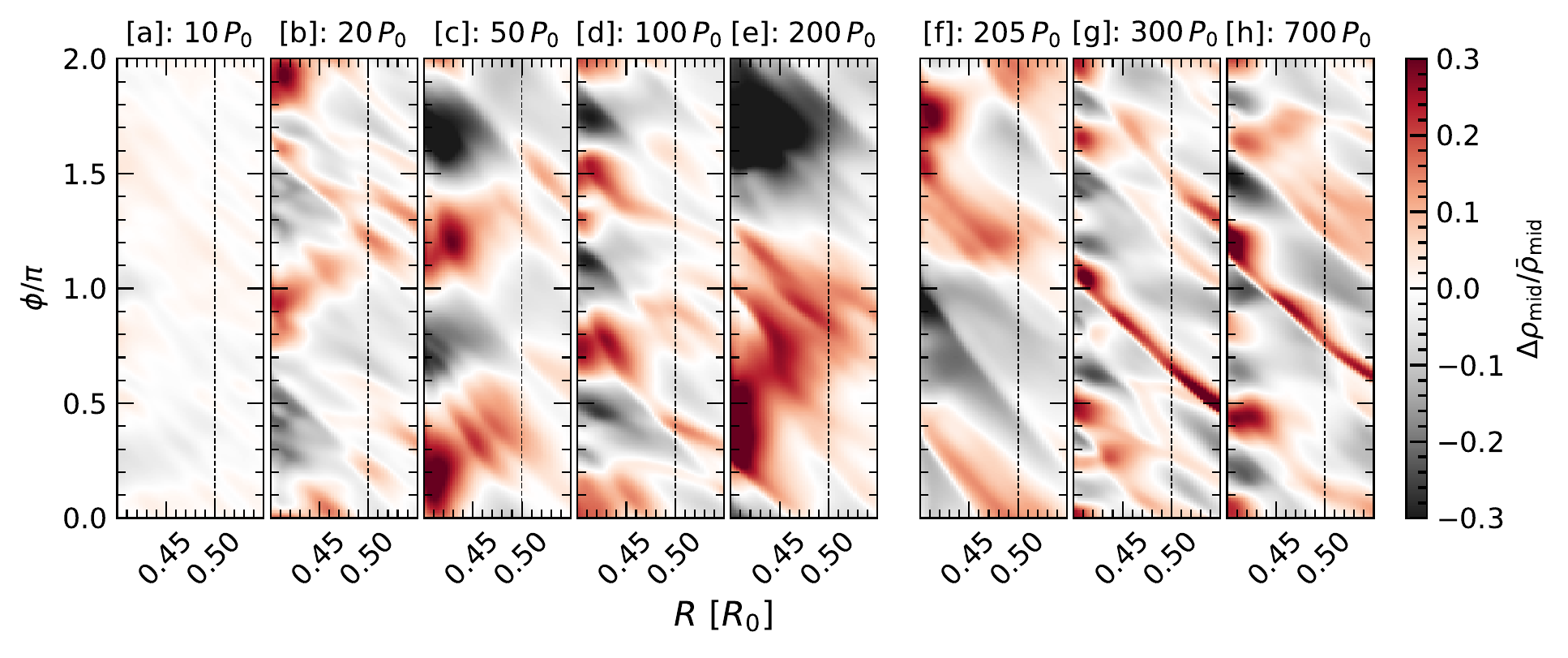}
	\parbox{\textwidth}{\caption{Time evolution of deviations in the midplane density $\rho_\text{mid}$ with respect to its azimuthal average, showing the development of over- and under-dense areas near the inner radial boundary edge.
	The time shown on each panel refers to the time elapsed since the disk has been expanded to 3D (see Fig.~\ref{fig:timeline}), and the planet grows between panels \emph{e}--\emph{g}.
	The nonaxisymmetric features grow over time in panels \emph{a}--\emph{e}, but are dampened within a few orbits after the planet is introduced in the domain (panel \emph{f}). Panel \emph{h} shows that these features no longer grow in the presence of the embedded planet. Vertical dotted lines mark the end of the wave-damping zone.}}
	\label{fig:rho-unhappiness}
\end{figure}

In our 3D runs shown in Sect.~\ref{sec:3d} we noted the presence of a parasitic instability in the inner radial boundary, likely owing to a problematic interface between the domain and the wave damping zone. As shown in Fig.~\ref{fig:rho-unhappiness}\textcolor{blue}{.1}, nonaxisymmetric features in the form of over- and under-densities form on the inner boundary edge at $R=0.4\,R_0$ and launch spirals akin to the vortices discussed in Sects~\ref{sec:3d}~and~\ref{sec:3d-vortex}. These spirals permeate the active domain, contributing slightly to the total Reynolds stress and likely inhibiting VSI activity to an extent.
This phenomenon has been seen before \citep[e.g.,][see Fig.~3 therein]{barraza-etal-2021}, and might relate to the weaker vertical velocities observed near the inner boundary for low-mass planets in \citet{stoll-etal-2017a}.

\begin{figure}[h]
	\centering
	\includegraphics[width=\textwidth]{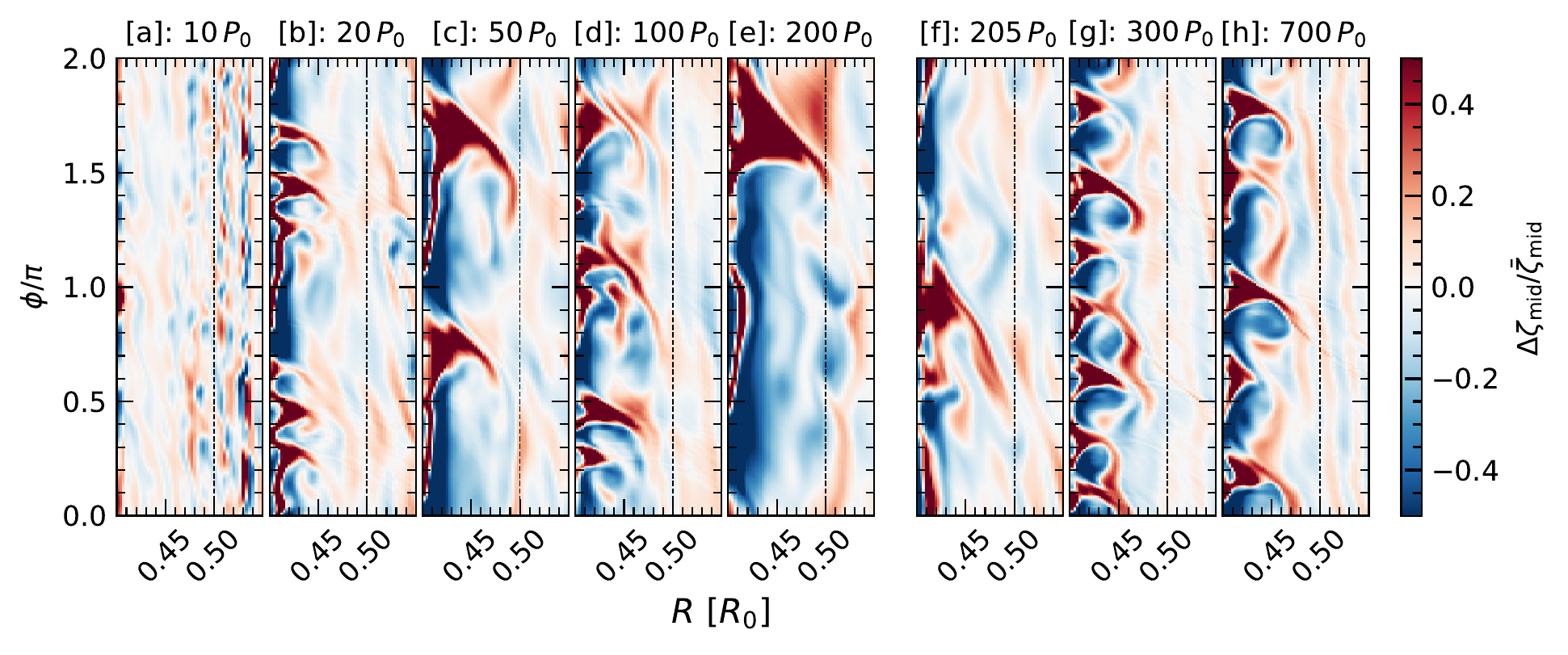}
	\parbox{\textwidth}{\caption{Time evolution of the deviations in the potential vorticity in the midplane according to Eq.~\eqref{eq:potential-vorticity} with respect to its azimuthal average. The figure is annotated similar to Fig.~\ref{fig:rho-unhappiness}\textcolor{blue}{.1}. In particular, panel \emph{a} shows that nonaxisymmetric features appear at the inner radial boundary edge at $R=0.4\,R_0$ as well as at the edge of the wave-damping zone at $0.5\,R_0$.}}
	\label{fig:vort-unhappiness}
\end{figure}

In an attempt to understand the origin of these nonaxisymmetric features, we calculate the potential vorticity at the midplane
\begin{equation}
	\label{eq:potential-vorticity}
	\zeta_\text{mid} = \frac{(\nabla\times\vel_\text{mid})\cdot\hat{z}}{\rho_\text{mid}}
\end{equation}
and plot it for various timestamps in Fig.~\ref{fig:vort-unhappiness}\textcolor{blue}{.2}. We find that vortex-like features start forming as early as 10\,$P_0$ into the simulation at both the inner boundary at $R=0.4\,R_0$ but also on the interface between the damping zone and the active domain at $0.5\,R_0$. While these features grow in size in the absence of a planet (see panels \emph{a}--\emph{e} in Figs.~\ref{fig:rho-unhappiness}\textcolor{blue}{.1}~\&~\ref{fig:vort-unhappiness}\textcolor{blue}{.2}),
the planet's inner spiral arms break up these overdensities, restabilizing the inner boundary to an extent and weakening the spiral arms they launch in the process (see panels \emph{f}--\emph{h} in the same figures). This would then explain---in part---why the VSI can remain active closer to the inner radial boundary in both our runs and in those by \citet{stoll-etal-2017a}.

\vfill

\end{appendix}

\end{document}